\documentclass[
,secnumarabic,showpacs
,amssymb,nobibnotes, aps, prd]{revtex4}
\usepackage{amsmath}
\usepackage{graphicx}
\input{epsf}
\newcommand{\be}{\begin{equation}}    
\newcommand{\ee}{\end{equation}}
\newcommand{\beq}{\begin{eqnarray}}
\newcommand{\eeq}{\end{eqnarray}}
\newcommand{\beqn}{\begin{eqnarray*}}
\newcommand{\eeqn}{\end{eqnarray*}}

\def\nn{\nonumber}

\def\ii{{\rm i}}   

\def\IL{\relax{\rm I\kern-.18em L}}

\def\cS{{\cal S}} 

\begin{document}


\title{Gravitational energy loss in high energy particle collisions:
ultrarelativistic plunge into a multidimensional black hole}
\author{Emanuele Berti}
    \altaffiliation{Email: berti@astro.auth.gr}
\affiliation{Department of Physics, Aristotle University
of Thessaloniki, Thessaloniki 54124, Greece}
\affiliation{McDonnell Center for the Space Sciences, Department of Physics,
Washington University, St. Louis, Missouri 63130, USA
\footnote{Present address: Groupe de Cosmologie et Gravitation
(GR$\epsilon$CO), Institut d'Astrophysique de Paris (CNRS), $98^{\rm bis}$
Boulevard Arago, 75014 Paris, France}}
\author{Marco Cavagli\`a}
    \altaffiliation{Email: cavaglia@phy.olemiss.edu}
\affiliation{Institute of Cosmology and Gravitation, University of
Portsmouth, Portsmouth PO1 2EG, UK}
\affiliation{Department of Physics and Astronomy, The University of Mississippi,
PO Box 1848, University, Mississippi 38677-1848, USA}
\author{Leonardo Gualtieri}
\altaffiliation{Email: leonardo.gualtieri@roma1.infn.it}
\affiliation{Dipartimento di fisica ``G.~Marconi'', Universit\`a di Roma ``La
Sapienza'' and Sezione INFN Roma 1, piazzale Aldo Moro 2, I-00185 Roma, Italy}
\date{\today}
\begin{abstract}
We investigate the gravitational energy emission of an ultrarelativistic
particle radially falling into a $D$-dimensional black hole. We numerically
integrate the equations describing black hole gravitational perturbations and
obtain energy spectra, total energy and angular distribution of the emitted
gravitational radiation. The black hole quasinormal modes for scalar, vector,
and tensor perturbations are computed in the WKB approximation. We discuss our
results in the context of black hole production at the TeV scale.
\end{abstract}
\maketitle

\section{Introduction}
Brane-world models describe the visible universe as a four-dimensional brane
embedded in a higher-dimensional bulk \cite{extradim}. A generic consequence of
the brane-world scenario is that the fundamental gravitational scale is lower
than the observed Planck scale. In some models, the fundamental scale is
lowered to values that would be accessible to next-generation particle
colliders, thus enabling laboratory-based studies of strong gravitational
physics via perturbative \cite{pertextradim} and nonperturbative events
\cite{nonpertextradim}. Ultrahigh energy cosmic rays could also probe
trans-Planckian energies \cite{uhecr}. The possibility that strong
gravitational effects such as black hole (BH) and brane formation could be
observed in the near future has sparked a lot of interest in the investigation
of nonperturbative gravitational phenomena in hard-scattering events
\cite{hardscat}. (For a review and more references, see
Ref.~\cite{Cavaglia:2002si}).

Trans-Planckian BH formation at energy scales much larger than the fundamental
gravitational scale is a classical process \cite{nonpertextradim}. The event is
dominated by the $s$-channel and the initial state is modelled by two classical
shock waves with given impact parameter. In this context, a major issue is the
estimate of the collisional energy loss. The hoop conjecture states that the
collision of two particles $ij$ with center-of-mass (c.m.)~energy $E_{\rm cm}$
and impact parameter smaller than the Schwarzschild radius $r_s(E_{\rm cm})$
forms a trapped surface \cite{hoop}. This event is formally described by the
process $ij\to {\rm BH}+E(X)$, where $E(X)$ denotes the collisional ``junk''
energy which does not contribute to the BH mass. The junk energy includes a
bulk component of gravitational radiation and other possible non-standard model
gauge fields, and a brane component of standard model collisional by-products
carrying the charge of the initial particles. The newly formed BH is expected
to decay first by loss of gauge radiation into the bulk and then by thermal
Hawking emission. The Hawking evaporation ends when the mass of the BH
approaches the fundamental gravitational scale. At this stage the BH either
decays completely by emitting the residual Planckian energy or leaves a stable
remnant with mass about the Planck mass \cite{relic}. Most of the observable
signatures of BH formation come from Hawking's phase and strongly depend on the
initial BH mass \cite{Cavaglia:2003hg}. Hence, a precise calculation of the
collisional energy loss is essential to the phenomenology of BH formation. 

A numerical estimate of the total collisional energy loss for spherically
symmetric BHs in $D\ge 4$ dimensions has been given by Yoshino and Nambu (YN)
\cite{Yoshino:2002tx} (see also Ref.~\cite{Vasilenko:2003ak}). The YN approach
evaluates the total junk energy $E(X)$ by investigating the formation of the BH
apparent horizon \cite{apphor}. The colliding particles are assumed massless,
uncharged and pointlike. Each particle is modelled by an infinitely boosted
Schwarzschild solution with fixed energy. This solution describes a
plane-fronted gravitational shock wave corresponding to the Lorentz-contracted
longitudinal gravitational field (Aichelburg-Sexl wave)
\cite{Aichelburg:1970dh}. The collision is simulated by combining two shock
waves travelling in opposite directions. The apparent horizon arises in the
union of the two shock waves. The junk energy is estimated by comparing the
initial c.m.\ energy to the BH mass. The result is that the collisional energy
loss depends on the impact parameter and increases as the number of spacetime
dimensions increases.  

The YN method allows estimation of the total junk energy in the classical
uncharged point-particle approximation. However, it cannot  discriminate
between different components of $E(X)$, which is theoretically and
experimentally most important. In a realistic BH event such as a proton-proton
collision at LHC \cite{LHC}, the BH is formed by the collision of two partons.
The bulk component of the junk energy is dominated by gravitational radiation
and is invisible to the detector. The gravitational junk energy and the
invisible component of Hawking emission (neutrinos, gravitons \dots) add to the
total missing energy of the process. Therefore, the knowledge of collisional
energy loss in gravitational emission should provide a good estimate of the
different sources of energy loss and missing energy.

An accurate estimate of the gravitational collisional energy loss
would require the use of the full non-linear Einstein equations in $D$
dimensions. This is a formidable task, even in four
dimensions. Recently, significant advances in numerical relativity
allowed stable numerical simulations of BH-BH collisions for initial
BH separation of a few Schwarzschild radii in the non-linear
Einstein theory. The gravitational waveforms predicted by these
simulations are in excellent agreement with analytical results from
first and second order perturbation theory
\cite{Gleiser:1998rw}. Since the linearization of the Einstein equations
yields results which are surprisingly close to the full theory (see, e.g.,
Ref.~\cite{Anninos:1998wt}), BH perturbation theory is likely
to provide accurate estimates of gravitational wave emission in
higher-dimensional spacetimes.  Relying on this result, we compute the
gravitational wave emission in higher dimensions via a perturbative
approach. Our computation is the first of this kind to our knowledge.

The formalism for the computation of gravitational wave emission from
perturbed BHs was developed by Regge and Wheeler \cite{RW} and Zerilli
\cite{Zerilli}, who reduced the problem to the solution of two
Schr\"odinger-like equations. Davis {\it et al.} \cite{DRPP} computed
the energy radiated in the radial infall of a particle of mass $m_0$
starting from rest at infinity into a four-dimensional BH of mass
$M_{\rm BH}\gg m_0$. This study was later generalized to the radial
infall of a particle with finite initial velocity or starting at
finite distance from the BH \cite{Ruffini}. (For a more comprehensive
introduction to BH perturbation theory see, e.g.,
Refs.~\cite{perturb}). Cardoso and Lemos \cite{Cardoso:2002ay,
Cardoso:2002yj} have recently investigated the plunge of
ultrarelativistic test particles into a four-dimensional static BH and
along the rotation axis of a Kerr BH, improving early estimates by
Smarr \cite{Smarr:1977}. In this paper we generalize these results to
higher dimensions by computing the gravitational radiation emitted by
an ultrarelativistic particle falling into a $D$-dimensional
spherically symmetric BH. Since wave propagation in odd-dimensional
curved spacetimes is not yet fully understood, we restrict our
investigation to even dimensions. (Wave late-time behavior and
propagation are very different in odd- and even-dimensional spacetimes
\cite{Cardoso:2002pa, Cardoso:2003jf, Barvinsky:2003jf}. Moreover,
open issues in the definition of asymptotic flatness
\cite{Hollands:2003ie} do not allow an unambiguous definition of
``gravitational waves radiated at infinity'' in odd dimensions.)

We model the particle collision as a relativistic test particle
plunging into a BH with mass $M_{\rm BH}=E_{\rm cm}$. We use recent
results of $D$-dimensional gravitational-wave theory by Cardoso {\it
et al.} \cite{Cardoso:2002pa} and the $D$-dimensional extension of
Zerilli's formalism by Kodama and Ishibashi (KI) \cite{Kodama:2000fa,
Kodama:2003jz, Ishibashi:2003ap, Kodama:2003kk}, which reduces the
problem to the solution of three Schr\"odinger-like equations. Our
method provides a simple and relativistically consistent estimate of
the collisional gravitational emission in higher dimensions. We derive
the emitted energy in terms of the wave amplitude and study the
angular dependence of the radiation using the KI formalism. We also
present a systematic calculation of BH quasinormal modes (QNMs) for
the different perturbations in the WKB approach, extending recent
calculations by Konoplya \cite{Konoplya:2003ii, Konoplya:2003dd}. We
show that there is a significant relation between the QNM frequencies
and the spectral content of the emitted radiation.

The outline of the paper is as follows. In Sec.~\ref{pe-qnms} we introduce our
notations and the basic equations. In Sec.~\ref{num} we briefly describe our
numerical approach to the computation of gravitational wave emission (details
are in the Appendices). Section \ref{res} contains the main results of the
paper. Conclusions are presented in Sec.~\ref{concl}.
\section{Perturbation equations and quasinormal modes\label{pe-qnms}}
In the next subsection we introduce the background metric and the KI
perturbation equations \cite{Kodama:2003jz}. In Subsec.~\ref{qnms} we
describe the method to compute the BH QNMs.
\subsection{Background metric and perturbation equations}
The spherically symmetric BH in $D=n+2$ dimensions is described by the
Schwarzschild-Tangherlini metric \cite{Tangherlini:1963}
\be\label{schw}
ds^2=-f(r)dt^2+\frac{dr^2}{f(r)}+r^2 d\Omega_n^2\,,
\ee
where $d\Omega_n$ is the metric of the $n$-dimensional unit sphere $S^n$, and 
\be\label{fdef}
f(r)=1-\frac{2M}{r^{n-1}}\,.
\ee
The BH mass $M_{\rm BH}$ is given in terms of the parameter $M$ by
\be\label{mass}
M_{\rm BH}=\frac{nM{\cal A}_n}{8\pi c^2 G_{n+2}}\,,
\ee
where ${\cal A}_n=2\pi^{(n+1)/2}/\Gamma[(n+1)/2]$ is the area of $S^n$,
$G_{n+2}$ is the $(n+2)$-dimensional Newton constant, and $c$ is the speed of
light. We will set $G_{n+2}=1$ and $c=1$ in the following. The
$(n+2)$-dimensional tortoise coordinate $r_*$ is defined by
\be\label{tort1}
\frac{dr_*}{dr}=\frac{1}{f(r)}\,.
\ee
Integrating Eq.~(\ref{tort1}) we find 
\be
r_*=r+\frac{2M}{n-1}\sum_{j=0}^{n-2}\frac{\ln(r/\alpha_j-
1)}{\alpha_j^{n-2}}\,,
\ee
where
\be
\alpha_j=(2M)^{1/(n-1)}{\rm e}^{2\pi{\rm i}j/(n-1)},\qquad
(j=0,\dots,n-2)\,,
\ee
and the integration constant has been chosen to make the argument of the
logarithm dimensionless. Here and throughout the paper we use the notations of
Refs.~\cite{Kodama:2003jz,Kodama:2000fa}; the indices $(\mu,\nu)$, $(i,j)$, and
$(a,b)$ denote the coordinates of the $D$-dimensional spacetime, the
coordinates of $S^n$, and the coordinates of the two-dimensional spacetime
$(t,r)$, respectively.

Kodama and Ishibashi \cite{Kodama:2003jz} showed that the
gravitational perturbation equations for this metric can be reduced to
Schr\"odinger--like wave equations:
\be\label{perteq}
\left(\frac{d^2}{dr_*^2}+\omega^2-V\right)\Phi=0\,,
\ee
where the potential $V$ depends on the kind of perturbation.  Setting
$x\equiv 2M/r^{n-1}$, the potential for scalar perturbations is
\be\label{pscalar}
V_S=\frac{f(r)Q(r)}{16r^2H^2}\,,
\ee
where
\be
\kappa^2=l(l+n-1),\qquad l=0,~1,~2,\dots\qquad
m=\kappa^2-n\,,\qquad
H(r)=m+\frac{n(n+1)}{2}x\,,
\ee
and
\beq
Q(r)&=&n^4(n+1)^2x^3+n(n+1)[4(2n^2-3n+4)m+n(n-2)(n-4)(n+1)]x^2\nn\\
&-&12n[(n-4)m+n(n+1)(n-2)]mx+16m^3+4n(n+2)m^2.
\eeq
Equation (\ref{pscalar}) reduces to the Zerilli equation \cite{Zerilli} for 
$n=2$. The potential for vector perturbations is
\be\label{pvector}
V_V=\frac{f}{r^2}\left[\kappa_V^2+1+\frac{n(n-2)}{4}-
\frac{3n^2M}{2r^{n-1}}\right]\,,
\ee
where
\be
\kappa_V^2=l(l+n-1)-1\,,\qquad l=1,~2,\dots\,.\\
\ee
Equation (\ref{pvector}) reduces to the Regge-Wheeler equation \cite{RW} for 
$n=2$. Finally, the potential for tensor perturbations is
\be\label{ptensor}
V_T=\frac{f}{r^2}\left[\kappa_T^2+2+\frac{n(n-2)}{4}+\frac{n^2M}
{2r^{n-1}}\right]\,,
\ee
where
\be
\kappa_T^2=l(l+n-1)-2,\qquad l=1,~2,\dots\,.\\
\ee
Equation (\ref{ptensor}) was derived by Gibbons and Hartnoll
\cite{Gibbons:2002pq} in a more general case (see also \cite{Neupane},
where a Gauss-Bonnet term is included) and has no equivalent in four
dimensions.
\subsection{Quasinormal modes\label{qnms}}
The knowledge of the QNM frequencies of multidimensional BHs enables a
clear physical interpretation of their gravitational emission. QNMs
are free damped BH oscillations which are characterized by pure
ingoing radiation at the BH horizon and pure outgoing radiation at
infinity. The no-hair theorem implies that QNM frequencies depend only
on the BH mass, charge and angular momentum.  Numerical simulations of
BH collapse and BH-BH collision show that, after a transient phase
depending on the details of the process, the newly formed BH has a
{\it ringdown} phase, i.e., it undergoes damped oscillations that can
be described as a superposition of slowly damped QNMs (modes with
small imaginary part). Furthermore, the QNMs determine the late-time
evolution of perturbation fields in the BH exterior (for comprehensive
reviews on QNMs see Refs.~\cite{KS}).

Gravitational radiation from four-dimensional astrophysical BHs is
dominated by slowly damped modes. In the following we show that these
also dominate the emission of gravitational radiation in higher
dimensions and determine important properties of the energy
spectra. Recently, Konoplya computed slowly damped QNMs of
higher-dimensional BHs \cite{Konoplya:2003ii, Konoplya:2003dd} using
the WKB method. This method is known to be inaccurate for large
imaginary parts, but it is accurate enough for the slowly damped modes
which are relevant in our context. Therefore, QNM frequencies for
scalar, vector, and tensor gravitational perturbations are computed
here in the WKB approximation. Our results are in good agreement with
those presented by Konoplya in Ref.~\cite{Konoplya:2003dd} (modulo a
different normalization). At variance with
Ref. \cite{Konoplya:2003dd}, we concentrated on uncharged black holes
in asymptotically flat, even dimensional spacetimes. We extended
Konoplya's calculation in two ways: i) in addition to the fundamental
QNM we also computed the first two overtones; ii) we carried out our
calculations for a much larger range of values of $l$
(Ref. \cite{Konoplya:2003dd} only shows results for $l=2$ and $l=3$).

The method consists in applying the WKB approximation to the potential in
Eq.~(\ref{perteq}) with appropriate boundary conditions. The result is a pair
of connection formulae which relate the amplitudes of the waves on either side
of the potential barrier, and ultimately yield an analytical formula for the
QNM frequencies (for details see Refs.~\cite{Iyer:np, Iyer:nq}). The WKB QNM
frequencies $\omega^2$ are given in terms of the potential maximum $V_0$ and of
the potential derivatives at the maximum by
\be\label{WKBf}
\omega^2=\left(V_0+\sqrt{-2V_0''}\Lambda\right)-i\left(j+\frac{1}{2}\right)
\sqrt{-2V_0''}\left(1+\Omega\right)\,,\qquad j=0,1,2,\dots\,,
\ee
where
\beq
\Lambda&=&\frac{1}{\sqrt{-2V_0''}}\left[\frac{1}{8}\frac{~V_0^{(4)}}
{V_0''}\left(\frac{1}{4}+\alpha^2\right)-\frac{1}{288}\left(\frac{V_0'''}
{V_0''}\right)^2
\left(7+60\alpha^2\right)\right]\,,\\
\Omega&=&\frac{1}{\sqrt{-2V_0''}}\left[
\frac{5}{6912}\left(\frac{V_0'''}
{V_0''}\right)^4\left(77+188\alpha^2\right)-
\frac{1}{384}\left(\frac{V_0'''{}^2V_0^{(4)}}
{V_0''{}^3}\right)\left(51+100\alpha^2\right)+\right.\\
&&~~~~~~~~~~\left.+\frac{1}{2304}\left(\frac{~V_0^{(4)}}
{V_0''}\right)^2\left(67+68\alpha^2\right)+
\frac{1}{288}\left(\frac{V_0'''V_0^{(5)}}
{V_0''{}^2}\right)\left(19+28\alpha^2\right)-
\frac{1}{288}\left(\frac{~V_0^{(6)}}
{V_0''}\right)\left(5+4\alpha^2\right)
\right]\,,\nn
\eeq
$\alpha=j+1/2$ and $j$ is the mode index. The QNM frequencies for the
scalar, vector, and tensor potentials of Sec.~\ref{pe-qnms} and
various dimensions are shown in Tables \ref{QNM2}-\ref{QNM8} and will
be discussed in Sec.~\ref{res}. Let us stress that the application
of the WKB technique is questionable in a few higher-dimensional
cases; for $l=2$ and $l=3$ the vector and scalar potentials in $D>6$
are not positive definite and/or display a second, small scattering
peak close to the BH horizon. An accurate analysis of these potentials
would require a refinement of the standard WKB technique, which is not
presented here. These special cases are denoted with italic numbers in
Tables \ref{QNM6}-\ref{QNM8}.

We mention that highly damped QNMs of four- and higher-dimensional BHs
have recently become a subject of great interest in a different
context. A few years ago, Hod proposed to use Bohr's correspondence
principle to determine the BH area quantum from highly damped BH QNMs
\cite{Hod:1998vk}. Hod's proposal is quite general: the asymptotic QNM
frequency for scalar perturbations of a non-rotating BH in $D$
dimensions is the same as in four dimensions \cite{Motl:2003cd}. Quite
notably, this result holds also for scalar, vector and tensor
gravitational perturbations \cite{Birmingham:2003rf,Motl:2003cd}.
Ref.~\cite{AreaQuantum} contains a partial list of references on
recent developments in this field.
\section{Integration method\label{num}}
The computation of the gravitational wave emission of an
ultrarelativistic particle plunging into a BH requires the numerical
integration of the inhomogeneous wave equation for scalar
gravitational perturbations. (Vector and tensor gravitational
perturbations are not excited by a particle in radial infall.) The
source term $S^{(n)}$ for the corresponding wave equation in $n+2$
dimensions can be calculated from the stress-energy tensor of the
infalling particle. Details of the derivation are in Appendix
\ref{sourcet}.

The integration in $(n+2)$ dimensions proceeds as in four dimensions
\cite{DRPP, Ruffini}. A good summary of the integration procedure can be found
in Ref.~\cite{Cardoso:2002ay}. In this section we simply stress the differences
between the four- and the $(n+2)$-dimensional cases. For the sake of
simplicity, in our numerical integrations we set the horizon radius
$r_h=(2M)^{1/(n-1)}=1$. The equation for the scalar perturbations is
\be\label{inhom}
\left(\frac{d^2}{dr_*^2}+\omega^2-V_S\right)\Phi=S^{(n)}\,.
\ee
The general solution of Eq.~(\ref{inhom}) is obtained via a Green function
technique as follows. Consider two independent ({\it left} and {\it right})
solutions of the homogeneous equation with boundary conditions $\Phi_L\sim
e^{-\ii \omega r_*}$ for $r_*\to -\infty$, and $\Phi_R\sim e^{\ii \omega r_*}$
for $r_*\to +\infty$. For $r_*\to +\infty$ the left solution is a
superposition of ingoing and outgoing waves of the form
\be\label{infty}
\Phi_L\sim B(\omega)e^{\ii \omega r_*}+C(\omega)e^{-\ii \omega r_*}\,.
\ee
The Wronskian is given by $W=2\ii \omega C(\omega)$. The wave amplitude is
obtained from a convolution of the left solution with the source term
\be\label{inhomsol}
\Phi=\frac{1}{W}\int_{-\infty}^{+\infty}\Phi_L~S^{(n)}dr_*\,.
\ee
The energy spectrum can be expressed in terms of the wave amplitude as
(details of the derivation are given in Appendix \ref{app2})
\be\label{energy}
\frac{dE}{d\omega}=\frac{\omega^2}{16\pi}
\frac{n-1}{n}\kappa^2(\kappa^2-n)|\Phi|^2\,,
\ee
where $\kappa^2\equiv l(l+n-1)$. The Wronskian for a given value of
$\omega$ is obtained by integrating the homogeneous equation from a
point located as close as possible to the horizon, and expanding
$\Phi_L$ as
\be
\Phi_L\sim e^{-i\omega r_*}\left[1+a_{n+2}(r-1)+\dots\right]\,,
\ee
where
\be
a_{n+2}=\frac{
-(l^4+2l^3-l^2-2l+3)+(n-2)
\left[-2l^3+l^2+(n^2+1)l-(n^3+4n^2+n+6)/4\right]}
{(2\ii \omega-1)(l^2+l+1)+(n-2)
\left[-l^2+(2\ii \omega-n)l+(n+1)\ii \omega-(n^2+1)/2\right]
}\,.
\ee
$C(\omega)$ (and $W$) can be obtained with good accuracy by matching the
numerically integrated $\Phi_L$ to the asymptotic expansion
\be
\Phi_L\sim 
e^{\ii\omega r_*}\left[1+\frac{a_{n+2}(\omega)}{r}+
\frac{b_{n+2}(\omega)}{r^2}+\dots\right]+
e^{-\ii\omega r_*}\left[1+\frac{a_{n+2}(-\omega)}{r}+
\frac{b_{n+2}(-\omega)}{r^2}+\dots\right]\,,
\ee
where the leading-order coefficient is
\be
a_{n+2}(\omega)=\frac{\ii\left[
l^2+(n-1)l+n(n-2)/4
\right]}{2\omega}\,.
\ee
For given $n$, $l$ and $\omega$, the error on the Wronskian and on the
energy spectrum is typically of the order of $O(10^{-4})$.
\section{Results\label{res}}
The main results of our work are the computation of the QNM frequencies in the
WKB approximation, the computation of the energy spectra, and the estimate of
the total energy and angular distribution of the radiation emitted during the
plunge. These results are discussed in detail below.
\subsection{Quasinormal frequencies}
The WKB QNM frequencies for different even values of $n$ are listed in Tables
\ref{QNM2}-\ref{QNM8}. Each line shows the first three quasinormal frequencies
($j=0,~1,~2$) for scalar, vector, and tensor perturbations at given $l$. For
$n=2$ tensor perturbations do not exist. In this case the scalar and vector
entries correspond to the QNMs of the Zerilli and Regge-Wheeler equations,
which are known to be isospectral \cite{perturb}. The isospectrality is broken
for $n>2$. This has been shown analytically by Kodama and Ishibashi
\cite{Kodama:2003jz} and later verified numerically by Konoplya
\cite{Konoplya:2003dd}. The real and imaginary parts of scalar QNM frequencies
at given $n$, $l$ and $j$ are smaller than those of vector QNMs, which are in
turn smaller than those of tensor QNMs. Since scalar modes are the least
damped, they are likely to dominate the gravitational radiation emission.

As $l$ grows, the isospectrality tends to be restored. In the eikonal
limit $l\to \infty$ the centrifugal term of the potential dominates
and is the same for scalar, vector, and tensor perturbations. In this
limit, the QNM frequencies for all perturbations are
\be
\omega_R\sim \frac{n+2l-1}{2}
\left(\frac{2}{n+1}\right)^{\frac{1}{n-1}}\left(\frac{n-1}{n+1}\right)^{1/2},
\qquad
\omega_I\sim \frac{n-1}{2(n+1)^{1/2}}
\left(\frac{2}{n+1}\right)^{\frac{1}{n-1}}
(2j+1)\,.
\ee
The previous relation was derived in Ref.~\cite{Konoplya:2003ii} for
multidimensional BH perturbations induced by a scalar field. (Notice that the
normalization used in Ref.~\cite{Konoplya:2003ii} is different from ours.) Here
we have shown that it also holds for gravitational perturbations.
Isospectrality of scalar and gravitational perturbations is a common feature of
the eikonal limit and of the large-damping limit \cite{Birmingham:2003rf,
Motl:2003cd} for any $n$. 
\subsection{Multipolar components of the energy spectra}
\begin{figure}[h]
\centering
\includegraphics[angle=270,width=8cm,clip]{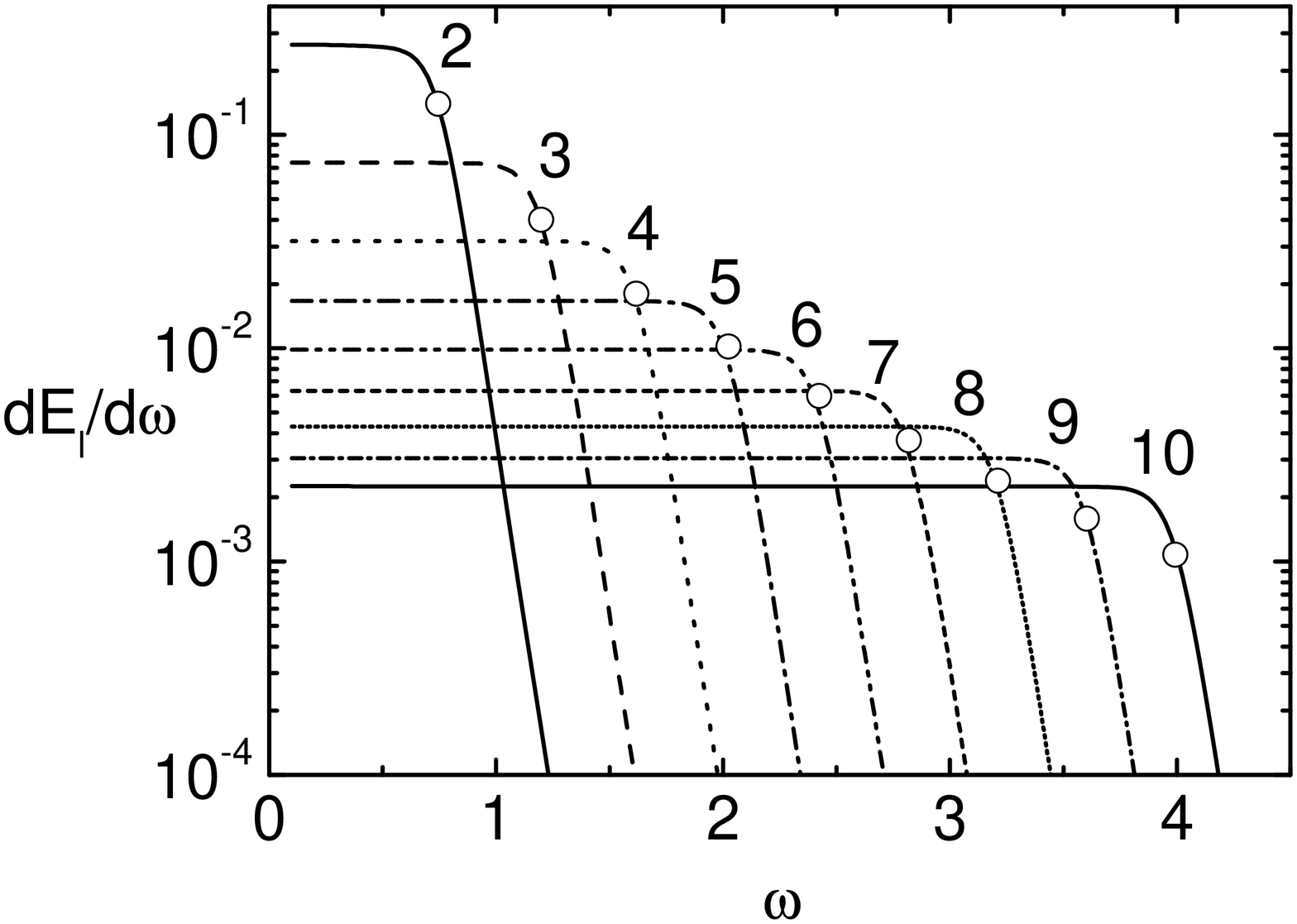}
\includegraphics[angle=270,width=8cm,clip]{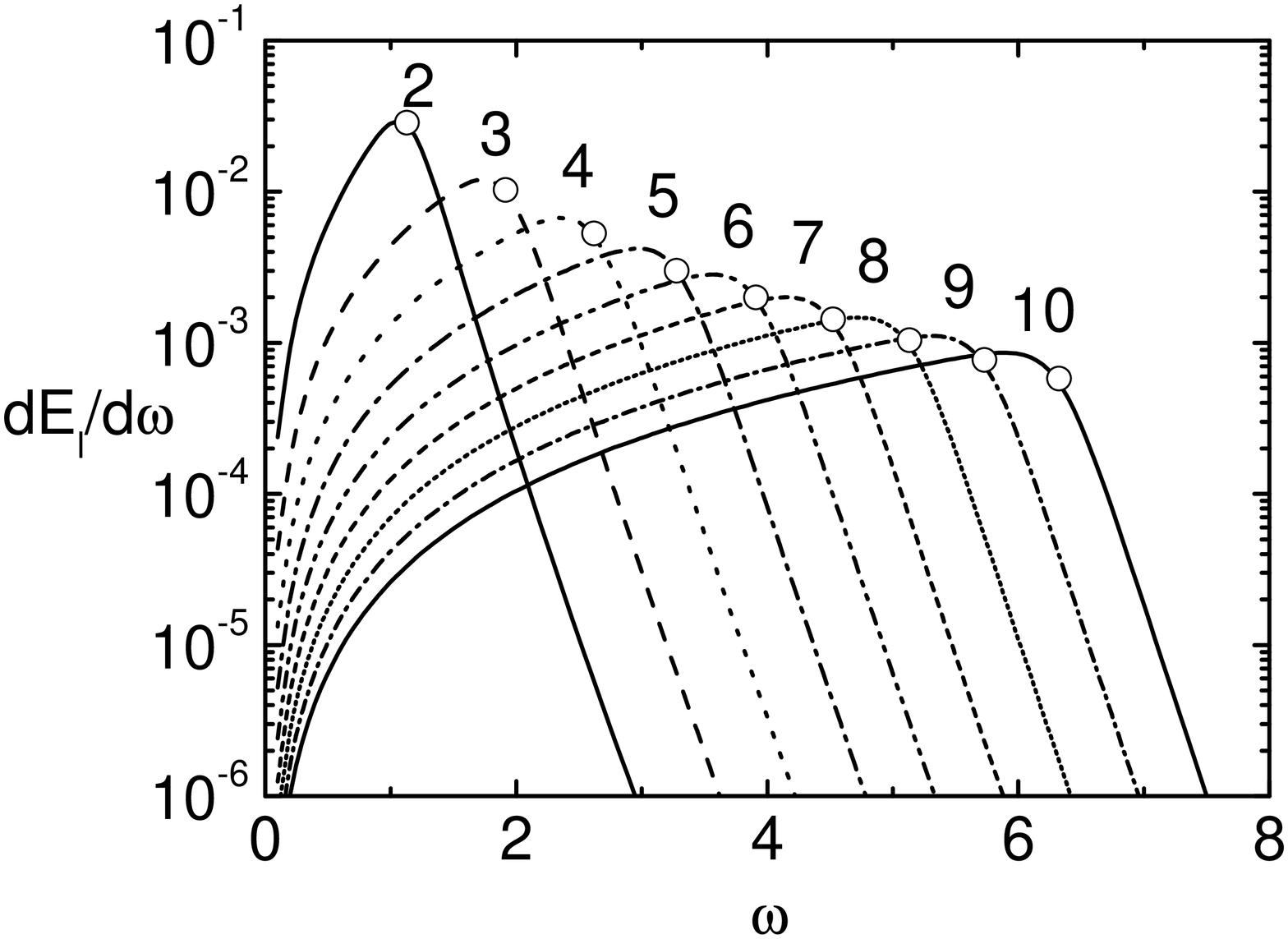}
\includegraphics[angle=270,width=8cm,clip]{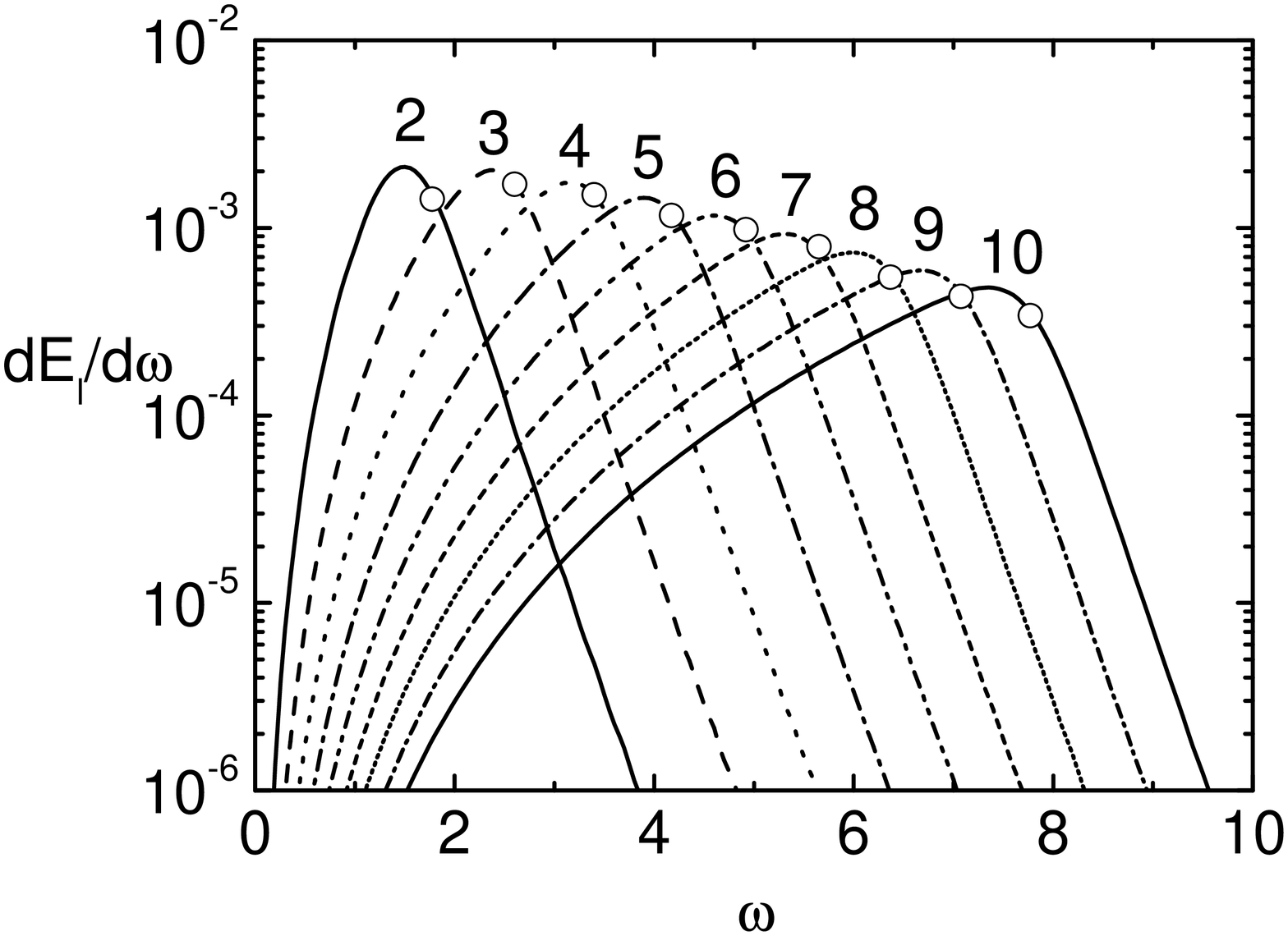}
\includegraphics[angle=270,width=8cm,clip]{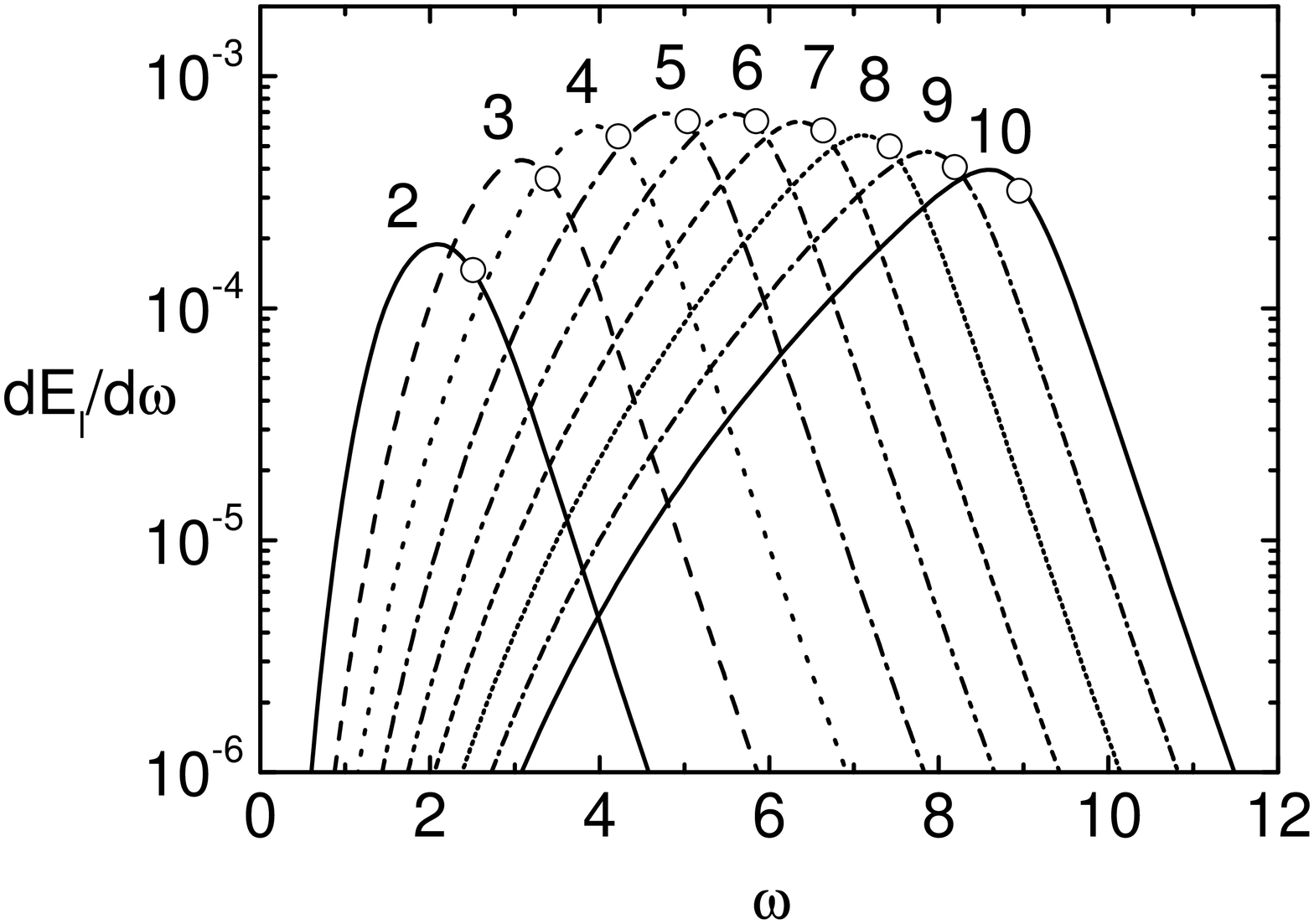}
\caption{
Multipolar components of the energy spectra up to $l=10$ for $n=2$,
$n=4$ (top left and top right panels), $n=6$ and $n=8$ (bottom left
and bottom right panels) in units $r_h=1$. Open circles mark the real
part of the fundamental scalar gravitational quasinormal frequency,
$\omega_{ln}$, for the given $l$ and $n$.
}
\label{fig1}
\end{figure}

\begin{figure}[h]
\centering
\includegraphics[angle=270,width=8cm,clip]{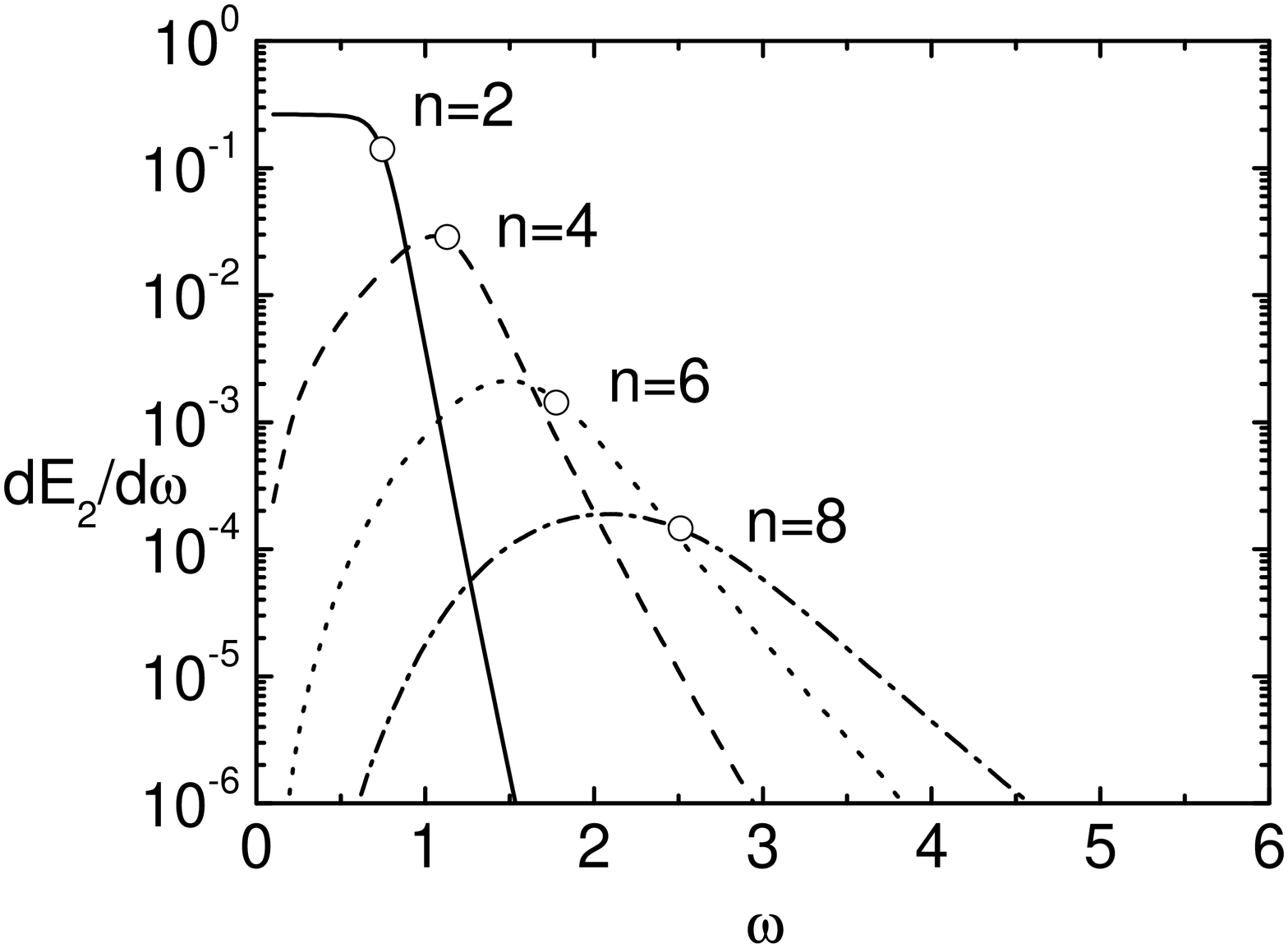}
\includegraphics[angle=270,width=8cm,clip]{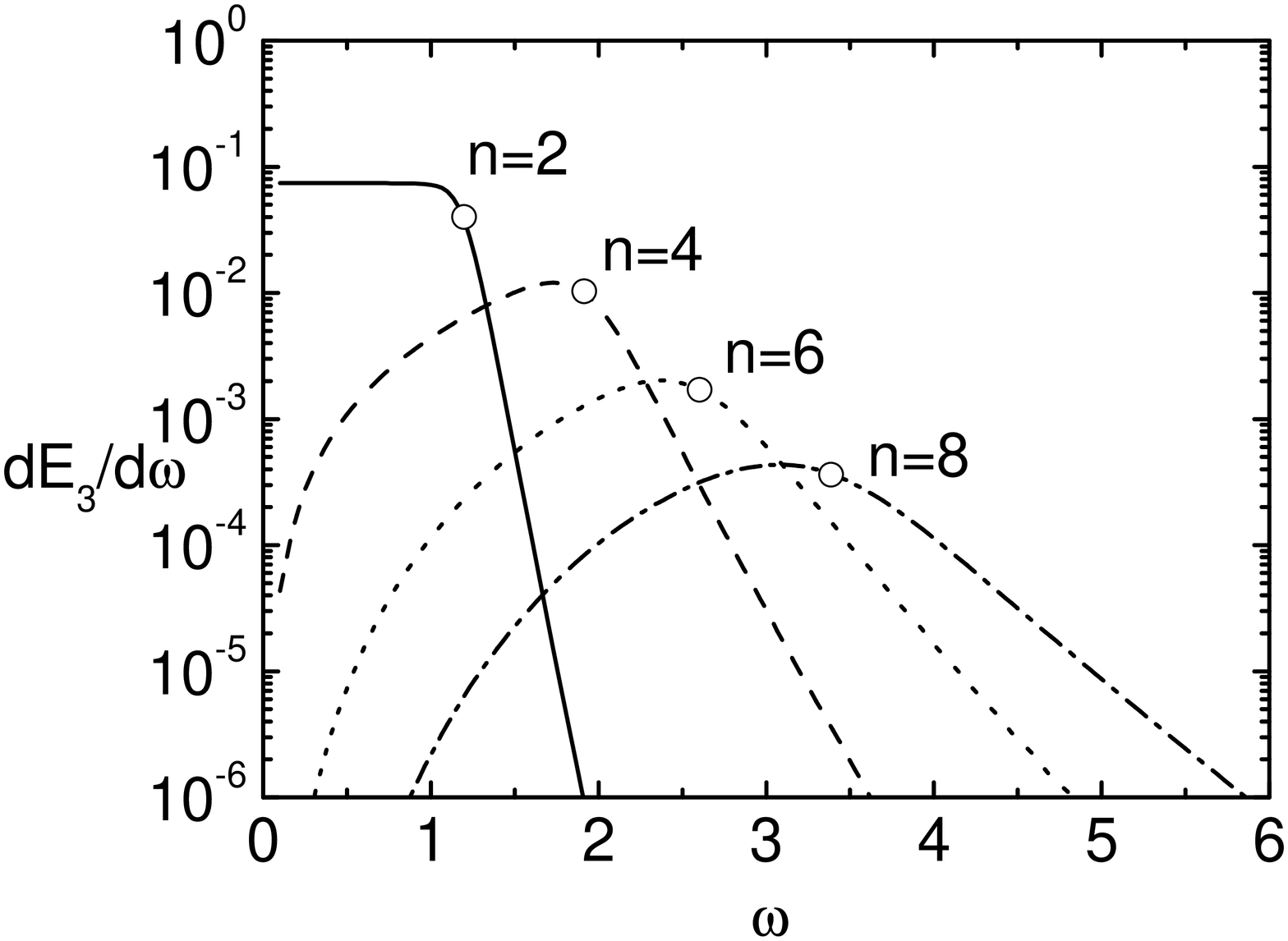}
\caption{
Multipolar components of the energy spectra at fixed $l$ for different
values of $n$ in units $r_h=1$. The left panel corresponds to $l=2$
and the right panel corresponds to $l=3$. Open circles mark the real
part of the fundamental scalar gravitational quasinormal frequency,
$\omega_{ln}$, for the given $l$ and $n$.
}
\label{fig2}
\end{figure}

The numerical integration described in Sec.~\ref{num} gives the energy spectra
of Figs. \ref{fig1} and \ref{fig2}. The spectra for $n=2$ (top left panel in
Fig. \ref{fig1}) are in excellent agreement with those of
Ref.~\cite{Cardoso:2002ay}.  The spectra are flat in the region between the
zero-frequency limit and a ``cutoff'' frequency $\omega_c$, beyond which they
fall exponentially to zero. The cutoff frequency $\omega_c$ is given by the
fundamental QNM frequency to a good level of accuracy. This result can be
understood in terms of gravitational-wave scattering from the potential barrier
which surrounds the black hole. $\omega^2$ plays the role of the energy in the
Schr\"odinger-like equation (\ref{perteq}). From Eq.~(\ref{WKBf}) it follows
that $\omega^2=V_0$ at first order in the WKB approximation. Therefore, only
the radiation with energy smaller than the peak of the potential is
backscattered to infinity; radiation with larger frequency is exponentially
suppressed.

The gravitational emission of a two-particle hard collision in higher
dimensions has been computed by Cardoso {\it et al.} \cite{Cardoso:2002pa}
using techniques developed in four dimensions by Weinberg \cite{Weinberg} and
later used by Smarr \cite{Smarr:1977}. The main result of
Ref.~\cite{Cardoso:2002pa} is that the spectra in $n+2$ dimensions grow as
$\omega^{n-2}$, thus the integrated spectra diverge as $\omega^{n-1}$.
Physically meaningful results for the total energy can only be obtained by
imposing some cutoff on the integrated spectra. Smarr \cite{Smarr:1977} first
suggested to use the inverse horizon radius as a cutoff. The relativistic
perturbative calculation in $n=2$ \cite{Cardoso:2002ay} shows that the cutoff
frequency at fixed $l$ is very close to the fundamental BH QNM. Therefore, the
cutoff frequency should be given by some ``weighted average'' of the
fundamental gravitational QNM frequencies \cite{Cardoso:2002pa}.

Our results for the spectra and the QNMs confirm the above picture.
Fig. \ref{fig1} shows that all spectra go to zero as $\omega\to 0$.
For $\omega<\omega_c$ the spectrum at fixed $l$ is
\be
\frac{dE_l}{d\omega}=f_{n,l}\omega^{n-2}\,,
\ee
where $f_{n,l}$ is a constant that can be found by a fit of the
spectra. For large $l$ $f_{n,l}$ decays as
\be
f_{n,l}=k_{n+2} l^{-3(n+2)/4}\,.
\ee
A fit of the numerical data gives $k_4=2.25$, $k_6=0.832$, $k_8=0.184$, and
$k_{10}=0.040$. Our result for $n=2$ is consistent with that of
Ref.~\cite{Cardoso:2002ay}.

As conjectured in Ref.~\cite{Cardoso:2002pa}, all spectra have a maximum at
some cutoff frequency $\omega_c$. This cutoff frequency is very close to the
fundamental QNM frequency $\omega_{ln}$ for (scalar) gravitational
perturbations with given $l$ and $n$, which is marked by open circles in Fig.
\ref{fig1} and Fig. \ref{fig2}. The deviation between $\omega_c$ and
$\omega_{ln}$ is of order 10 \% for low $l$, and decreases for large $l$
(compare Fig.~\ref{fig1} and Fig.~\ref{fig2} to the first column of Tables
\ref{QNM2}-\ref{QNM8}). The deviation is larger when the WKB method is least
reliable, namely for $l=2$ and $n>4$. In these cases, the location of the peaks
in the spectra can presumably be used as a more reliable estimate of the QNM
frequency. The spectrum decays exponentially for $\omega>\omega_c$ with an
$n$-dependent slope $\alpha_{n+2}$ (see Fig. \ref{fig2}):
\be
\frac{dE_l}{d\omega}\sim e^{-\alpha_{n+2} (\omega-\omega_c)}\,.
\ee
Thus the $\omega$-integrated multipolar contributions $\Delta E_l$ at given $l$
are finite. With our choice of units, Cardoso and Lemos \cite{Cardoso:2002ay}
find $\alpha_4=13.5 \alpha$ (here $\alpha$ is a constant of order unity which
cannot easily be determined because the spectra decay very quickly). Our
numerical fits give $\alpha_4\simeq 15$, in good agreement with their result.
In higher dimensions the constants $\alpha_{n+2}$ are comparatively easier to
determine. Their values are $\alpha_6\simeq 5.5$, $\alpha_8\simeq 3.4$, and
$\alpha_{10}\simeq 2.3$. It is not clear if there is any relation between this
$n$-dependent slope and the late-time tail behavior predicted in
Ref.~\cite{Cardoso:2003jf}.

\begin{figure}[h]
\centering
\includegraphics[angle=270,width=9cm,clip]{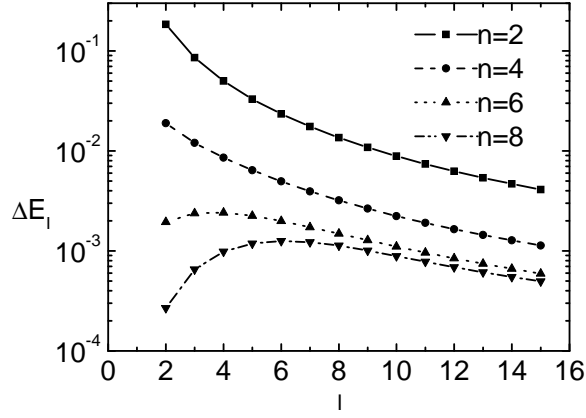}
\caption{
The integrated energy $\Delta E_l$ as a function of $l$ for different values of
$n$. The dominant multipolar component is $l=2$ only for $n<6$; this is
probably related to the appearance of a negative well in the scalar potentials
for $l=2$ and $n>4$. The dominant multipole is $l=4$ ($6$) for $n=6$  ($8$)
(see Table \ref{El}).
}
\label{fig3}
\end{figure}

Fig.~\ref{fig1} shows that higher multipoles contribute more as $n$ grows. 
This is evident when we look at the $\omega$-integrated multipolar components
of the energy spectra of Fig.~\ref{fig3}. The quadrupole ($l=2$) is dominant
only for $n=2$ and $n=4$. For $n=6$ and $n=8$ the dominant multipoles are $l=4$
and $l=6$, respectively (see Table \ref{El}). This effect may be related to the
appearance of a negative well in the scalar potentials for $l=2$ and $n>4$. It
would be interesting to understand better the physical relation between the
dominant multipole and the spacetime dimension.
\subsection{Total energy}
The total emitted energy is obtained by numerically integrating the results of
the previous section over $\omega$ and summing the multipolar components.
For large $l$ the integrated energy in the multipole $l$ can be fitted by
\be\label{Efit}
\Delta E_l=a_{n+2} l^{-b_{n+2}}\,,
\ee
where $(a_6=0.110, b_6=1.69)$, $(a_8=0.050, b_8=1.64)$, and $(a_{10}=0.022,
b_{10}=1.40)$ for $n=4$, $n=6$, and $n=8$, respectively. The coefficients
$(a_{n+2}, b_{n+2})$ have been obtained by fitting the data from $l=14$ to
$l=20$ and are weakly dependent on the chosen range of $l$. This variability
affects our final results on the total energy within less than a few percent.

Restoring the dependence on the BH horizon $r_h$ and on the conserved
particle energy $p_0$, the total emitted energy is
\be
E_{\rm em}=\frac{p_0^2}{M_{\rm BH}}\frac{n{\cal A}_n}{16\pi}
\sum_{l=2}^\infty \Delta E_l
\equiv
\frac{p_0^2}{M_{\rm BH}}\frac{n{\cal A}_n}{16\pi}
E^{(D)}_{\rm tot}\,
\equiv
\frac{p_0^2}{M_{\rm BH}}{\cal E}_{\rm tot}^{(D)}\,,
\ee
where $E^{(D)}_{\rm tot}$ is the ``dimensionless'' total energy
(expressed in the units $r_h=1$ that we used in our numerical
integrations).  We obtained the integrated spectra numerically up to
$l=20$ and extrapolated them for larger $l$ using the fits in Eq.\
(\ref{Efit}). Results are presented in Table \ref{Etot}.

Following Ref.~\cite{Cardoso:2002ay} we estimate the gravitational
energy loss for a collision of two particles with equal mass $M$ by
the replacement $p_0\to M$, $M_{\rm BH}\to M_{\rm tot}=2M$.  For $n=2$
this extrapolation gives results in good agreement with the
perturbative shock-wave calculation of Ref.~\cite{apphor} which
considers two BHs of equal mass. An analogous extrapolation for $n=2$
gives results in close agreement with the fully relativistic
computation \cite{Anninos:1998wt} for a particle starting from
infinity at rest. Therefore, we believe that our extrapolation should
provide a qualitative but realistic estimate. The results for
different values of $n$ are given in the last column of Table
\ref{Etot}. The gravitational energy loss is $\sim 13$\%, $\sim 10$\%,
$7$\%, and $8$\% for $n=2$ to $n=8$, respectively. The result for
$n=2$ is in good agreement with previous estimates \cite{apphor} (see
the discussion in Ref.~\cite{Cardoso:2002ay}).
\subsection{Angular dependence}
The angular dependence of the radiation is obtained by evaluating numerically
Eq.~(\ref{angular}) and Eq.~(\ref{angulartot}). Fig.~ \ref{fig4} shows the
angular dependence of the total energy up to $l=15$ for $n=2$, $n=4$ and $n=6$.

The angular distribution of the gravitational radiation in the BH
frame goes to zero along the axis of the collision ($\theta=0$, $\pi$)
in any dimensions.  Therefore, the gravitational emission is never
back or forward scattered. In four dimensions the angular spectrum of
the gravitational radiation increases rapidly at small $\theta$ and
becomes approximately flat at greater angles with a maximum in the
direction orthogonal to the axis of the collision, before falling
rapidly to zero for values of the angles close to $\pi$. The angular
distribution of the gravitational radiation for $n>2$ is peaked at
$\pm\theta$ and $\pi\pm\theta$, where $\theta$ is a small angle. The
difference between the behavior of the angular distribution in four
dimensions and in higher dimensions has no evident physical reason. It
would be interesting to further explore this point.
\section{Conclusion and perspectives\label{concl}}
In this paper we have computed the gravitational emission of a two-particle
collision in an even $D$-dimensional spacetime. We have presented the numerical
results for $D=4$ to $10$. The collision has been modelled as a massless test
particle plunging into a BH with mass equal to the c.m.\ energy of the event. 

According to our estimates, the total emitted energy in a head-on
collision with particles of equal mass ranges from $\sim 13$\% ($D=4$)
to $\sim 8$\% ($D=10$). This shows that the loss in gravitational
radiation is quite stable under variation of the spacetime dimension
and slightly decreases for higher $D$. The result for $D=4$ confirms
previous numerical and analytical calculations \cite{apphor}.

Our result contrasts with the YN estimation for the initial mass of a BH in
head-on collisions \cite{Yoshino:2002tx}. A possible explanation is that the
junk emission is not wholly gravitational emission. The YN method predicts the
mass within the apparent horizon to be $\sim 0.71 E_{\rm cm}$ in four
dimensions.  If all the junk energy were gravitational radiation, this would
amount to a total loss of around $30$\%. The disagreement is likely not due to
numerical uncertainties or inaccurate approximations: the YN mass decreases
for higher spacetime dimensions ($\sim 0.71 E_{\rm cm}$ to $\sim 0.58 E_{\rm
cm}$ for $D=4$ to $D=11$), whereas the loss in gravitational radiation remains
stable. Since both YN and our methods are purely gravitational, this ``dark
component'' of the junk radiation should describe the by-products of the
collision. According to this picture, $\sim 60$\% of the c.m.\ energy in ten
dimensions is trapped inside the horizon, $\sim 10$\% is emitted in
gravitational radiation, and $\sim 30$\% goes into particle by-products in the
final state. These could be the carriers of the initial charge in a collision
between charged particles. For $D>4$ a fraction of the by-products may be
emitted into the bulk.

Let us conclude by briefly discussing the phenomenological consequences of
these results for BH formation at the TeV scale. Although uncertainties may
affect the numerical estimates, different approaches now confirm that some of
the initial c.m.\ energy is not trapped inside the BH horizon. For head-on
collisions in $D=10$, for example, this junk energy ranges from $\sim 10$\%
(optimistic value -- our result) to $\sim 40$\% (pessimistic value -- YN
result). Hence, the initial mass of the BH formed in the collision could be
considerably smaller than the c.m.\ energy. The experimental signatures of BH
production at particle colliders and in ultrahigh energy cosmic ray events
strongly depend on the initial BH mass. The total multiplicity of the Hawking
phase in ten dimensions could be almost halved in the pessimistic case, leading
to a greater average energy of the emitted quanta. 

A thorough investigation of the effects of energy loss in TeV-scale BH
production is undoubtedly worth pursuing. Future research should focus on the
extension of the above results to spacetimes with odd dimensions and to
gravitational events with different geometries. BHs produced in colliders, for
instance, possess nonvanishing angular momentum. Rotating BHs are expected to
lose more energy in gravitational waves than Schwarzschild BHs of equal mass. A
larger gravitational emission is also expected for non-spherically symmetric
BHs. This is particularly relevant when the compactified space is asymmetric,
and some of the extra dimensions have size of order of the fundamental
gravitational scale. It would be extremely important to quantify these
differences. 
\acknowledgments
We warmly thank V.~Cardoso for his encouragement to carry
out this work and for several useful suggestions. M.~C. is grateful to
E.-J.~Ahn for many helpful comments on the content of this paper. This
work has been supported by the EU Programme `Improving the Human
Research Potential and the Socio-Economic Knowledge Base' (Research
Training Network Contract HPRN-CT-2000-00137). M.~C. is partially
supported by PPARC.
\appendix
\section{The source term\label{sourcet}}
In this Appendix we derive the source term of the KI equation that describes
the radial plunge of a massless particle into the $(n+2)$-dimensional BH.
The perturbation of the stress-energy tensor is
\be\label{T1}
\delta T^{\mu\nu}=-\frac{p_0}{\sqrt{-g}}\int d\lambda\delta^n(x-x(\lambda))
\frac{dx^{\mu}}{d\lambda}\frac{dx^{\nu}}{d\lambda}\,,
\ee
where $p_0$ is the conserved energy of the particle.
The only non-vanishing components of the particle velocity are $u^t$ and $u^r$.
Thus the source excites only scalar perturbations. Following the notations of
Ref.~\cite{Kodama:2000fa}, Eq.~(\ref{T1}) reads
\be\label{Tt}
\delta T_{\mu\nu}=\left(
\begin{array}{c|c}
\tau_{ab}\cS & 0 \\ \hline 0 & 0 \\
\end{array}
\right)\,,
\ee
where $\cS$ are the scalar harmonics and $\tau_{ab}$ are the non-vanishing
gauge-invariant perturbations of the stress-energy tensor. The BH+source system
is symmetric under rotation of the $(n-1)$-sphere $S^{n-1}$ \cite{Cardoso:2002cf}.
Consequently, the harmonic decomposition of the fields contains only 
harmonics invariant under $S^{n-1}$. We can write the metric of $S^n$ as
\be
d\Omega_{n}(\theta,\phi_1,\dots,\phi_{n-1})=
d\theta^2+\sin^2\theta d\Omega_{n-1}(\phi_1,\dots,\phi_{n-1})
\ee
and choose the trajectory of the test particle to be $\theta=0$. The harmonics
which are invariant under $S^{n-1}$ do not depend on $\phi_1,\dots,\phi_n$.
The scalar harmonics on $S^n$ belong to the representations $D^{(l)}$ of
$SO(n+1)$
\be
D^{(l)}=\underbrace{
\begin{array}{|c|c|c|c|}
\hline
&\dots&\\
\hline
\end{array}}_{l}\,.
\ee
Each harmonic is labelled by the index $l$ denoting its representation and by
additional indices in the representation. We fix a particular element of each
representation $D^{(l)}$ (the singlet under $SO(n-1)$) by requiring the
harmonics to be invariant under $S^{n-1}$. Therefore, the harmonics in the
expansion of the perturbations depend only on $l$ and on the dimension $n$ of
the sphere. A $(n+2)$-dimensional scalar field can then be expanded as
\cite{Cardoso:2002cf}
\be
\phi(t,r,\theta,\phi_1,\dots,\phi_{n-1})=\sum_l \tilde\phi_l(t,r)
\cS^{(nl)}(\theta)\,,
\ee
where $\cS^{(nl)}(\theta)$ satisfy
\be\label{eqGeg}
D_iD^i \cS^{(nl)}=-\kappa^2\cS^{(nl)},\qquad\kappa^2\equiv l(l+n-1)\,,
\ee
and
\be\label{normalizY}
\int d\Omega_n\cS^{(nl)}\cS^{*(nl')}=\delta_{ll'}\,.
\ee
The solution of Eq.~(\ref{eqGeg}) is
\be
\cS^{(nl)}(\theta)=K^{(nl)}C_l^{(n-1)/2}(\theta)\,,
\ee
where $C_l^{(n-1)/2}(\theta)$ are Gegenbauer polynomials \cite{Abr}
and $K^{(nl)}$ are normalization factors. Using Eq.~(\ref{normalizY})
we have
\be
K^{(nl)}=\left[\frac{2^{3-n}\pi^{n/2+1}}{\Gamma\left(\frac{n}{2}\right)}
\frac{\Gamma(l+n-1)}{\left(l+\frac{n}{2}-\frac{1}{2}\right)
\Gamma\left(\frac{n}{2}-\frac{1}{2}\right)^2\Gamma(l+1)}\right]^{-1/2}\,.
\ee
The scalar harmonics for the source are obtained setting $\theta=0$:
\be
S^{(nl)}(\theta=0)=K^{(nl)}\frac{(l+n-2)!}{(n-2)!l!}\,.
\ee
For a massive particle in radial geodesic motion
\be\label{dtdr}
\frac{dt}{dr}=-\frac{1}{f(r)}\,.
\ee
From equations (\ref{dtdr}) and (\ref{tort1}) it follows that
\be
r_*(r)=-t(r)\,,
\ee
where we have set to zero the integration constant. The  stress-energy tensor
perturbations are 
\beq\label{deltaT}
\delta T_{rr}(t,r,\theta,\phi_1,\dots)&=&-p_0 f^{-2}r^{-n}\delta(t-t(r))
\delta^n(\Omega_n)\,,\nn\\
\delta T_{tr}(t,r,\theta,\phi_1,\dots)&=&-p_0 f^{-1}r^{-n}\delta(t-t(r))
\delta^n(\Omega_n)\,.
\eeq
($\delta T_{tt}$ does not contribute to the source of the KI equation.)
Integrating Eqs.~(\ref{deltaT}) on $S^n$ and applying a
Fourier-transform, the gauge-invariant perturbations are
\beq
\tau_{rr}(\omega,r)&=&-p_0 f^{-2}r^{-n}\frac{e^{\ii\omega t(r)}}{\sqrt{2\pi}}
S^{(nl)}(\theta=0)\,,\nn\\
\tau_{tr}(\omega,r)&=&-p_0 f^{-1}r^{-n}\frac{e^{\ii\omega t(r)}}{\sqrt{2\pi}}
S^{(nl)}(\theta=0)\,.\label{tauexpr}
\eeq
The source term $S^{(n)}$ for scalar gravitational perturbations is
obtained in terms of $\tau_{ab}$ by substituting Eqs.~(\ref{tauexpr})
in Eq.~(5.44) of Ref.~\cite{Kodama:2003kk}, where
\be
S_{ab}=8\pi r^{n-2}\tau_{ab},\qquad
S_a=S_T={\cal A}={\tilde J}_a=0\,.
\ee
The result is:
\beq
\label{S2}
S^{(2)}
&=&e^{-\ii\omega r_*}
\frac{8\sqrt{4l+2}}{\ii\omega r}
\frac{(l-1)(l+2)(r-1)}{\left[(l+2)(l-1)r+3\right]^2}\,,\\
\label{S4}
S^{(4)}
&=&e^{-\ii\omega r_*}\frac{16\sqrt{(2l+3)\lambda_2}}
{\sqrt{2\pi}\ii\omega r^3}
\frac{\left[(l^2+3l)(r^6-r^3)+5-4r^6-r^3\right]}
{[(l+4)(l-1)r^3+10]^2}\,,\\
\label{S6}
S^{(6)}
&=&e^{-\ii\omega r_*}\frac{6\sqrt{(2l+5)\lambda_4}}
{\pi\ii\omega r^4}
\frac{\left[(l^2+5l)(r^{10}-r^5)+14-6r^{10}-8r^5\right]
}{[(l+6)(l-1)r^5+21]^2}\,,\\
\label{S8}
S^{(8)}
&=&e^{-\ii\omega r_*}\frac{4\sqrt{(2l+7)\lambda_6}}
{\sqrt{3\pi^3}\ii\omega r^5}
\frac{\left[(l^2+7l)(r^{14}-r^7)+27-8r^{14}-19r^7\right]}
{[(l+8)(l-1)r^7+36]^2}\,,
\eeq
where $\lambda_k\equiv[(l+k)(l+k-1)\dots(l+1)]$. The BH horizon $r_h$
and the particle energy $p_0$ have been set equal to one. Equation
(\ref{S2}) coincides with Eq.~(8) of Ref.~\cite{Cardoso:2002ay}.
\section{Energy and angular distribution}
\label{app2}
We derive here the formulae for the energy and the angular dependence
of the gravitational emission. Gravitational waves in a
$(n+2)$-dimensional spacetime behave asymptotically as $\sim~r^{-n/2}$
\cite{Cardoso:2002pa} and possess $n(n+1)/2-1$ degrees of freedom. The
transverse-traceless (TT) gauge is defined by $\delta g_{ab}=0$,
$\delta g_{ai}=0$, and $\delta g_{ij}\gamma^{ij}=0$. These conditions
can be chosen by imposing the harmonic gauge in the wave zone and
using the remaining gauge freedom to constrain $\delta g_{tt}=0$, $\delta
g_{ti}=0$, and $\gamma^{ij}\delta g_{ij}=0$.

We separate the angular part of the perturbations using tensor
spherical harmonics, following Ref. \cite{Kodama:2003jz}. In the
TT-gauge, the only nonvanishing term in the decomposition is the $H_T$
component.  Hence, the gauge invariant quantities for scalar
perturbations ($F$ and $F_{ab}$) depend only on $H_T$:
\be
F=\frac{1}{n}H_T+\frac{1}{r}(D^ar)X_a\,,\qquad
F_{ab}=D_aX_b+D_bX_a\,,
\ee
where
\be
X_a=\frac{r^2}{\kappa^2}D_aH_T\,.
\ee
The scalar perturbation $\Phi$ is written in terms of the gauge invariant
quantities as
\be
\Phi=\frac{n\tilde Z-r(X+Y)}{r^{n/2-1}[\kappa^2-n+n(n+1)x/2]}\,,
\ee
where
\be
X+Y=-2nr^{n-2}F\,,\qquad
\tilde Z=\frac{1}{{\rm i}\omega}r^{n-2}F^r_{~t}\,.
\ee
Setting
\be\label{ansatz1}
H_T\rightarrow\frac{A}{r^{n/2}}e^{\ii\omega r_*}\,,
\ee
the asymptotic behavior of $\Phi$ is
\be
\lim_{r\rightarrow\infty}\Phi=\frac{2A}{\kappa^2} e^{\ii\omega r_*}\,.
\ee
The asymptotic behavior of $h_{ij}$ in the TT-gauge is 
\be
h^{TT}_{ij}=2r^2H_T {\cal S}_{ij}\sim\frac{\kappa^2\Phi}{r^{n/2-2}}
{\cal S}_{ij}=
\frac{\Phi}{r^{n/2-2}}\left(D_iD_j {\cal S}+\frac{\kappa^2}{n}\gamma_{ij}
{\cal S}\right)\,.
\label{hijas}
\ee
The energy-momentum pseudotensor does not depend on the spacetime dimension and
is given by \cite{Cardoso:2002pa}
\be\label{em-tensor1}
\frac{dE}{dSdt}=<t^{00}>=\frac{\omega^2}{32\pi}<h^{TT}_{ij}h^{TT\,ij}>\,,
\ee
where $h^{TT}_{ij}$ are metric perturbations in the time domain. Using
Parseval's theorem, the energy-momentum pseudotensor in the frequency domain is
\be\label{em-tensor2}
\frac{dE}{dSd\omega}=\frac{\omega^2}{32\pi}
<\tilde h^{TT}_{ij}\tilde h^{*TT\,ij}>\,,
\ee
where $\tilde h^{TT}_{ij}$ are now the metric perturbations in the
frequency domain. Substituting Eq.~(\ref{hijas}) in
Eq.~(\ref{em-tensor2}) we get
\be
\frac{dE}{dSd\omega}=\frac{\omega^2}{32\pi}\frac{|\Phi|^2}{r^n}
\left(D_iD_j{\cal S}+\frac{\kappa^2}{n}\gamma_{ij}{\cal S}\right)
\left(D_kD_l{\cal S}^*+\frac{\kappa^2}{n}\gamma_{kl}{\cal S}^*\right)\gamma^{ik}\gamma^{jl}
\,.
\ee
Integrating on the sphere $dS=r^nd\Omega_n$ and using the relations 
\be
D_iD^i {\cal S}=-\kappa^2{\cal S}\,,\qquad
[D_j,D_i]V^j=R_{ki}V^k=(n-1)V_i\,,
\ee
where $V^i$ is a generic vector (here and in the following, indices are raised and lowered with
$\gamma^{ij}$), we find the ``two-sided'' power spectrum
\be\label{twosided}
\frac{dE}{d\omega}_{two-sided}=\int dS^{(n)}\frac{dE}{dSd\omega}
=\frac{\omega^2}{32\pi}
\frac{n-1}{n}\kappa^2(\kappa^2-n)|\Phi|^2\,.
\ee
The ``one-sided'' spectrum in Eq.~(\ref{energy}) is obtained by multiplying
Eq.~(\ref{twosided}) by two. In the four-dimensional limit the one-sided
spectrum is
\be\label{zerilli}
\left.\frac{dE}{d\omega}\right|_{n=2}=\frac{\omega^2}{32\pi}
\frac{(l+2)!}{(l-2)!}|\Phi|^2\,.
\ee
Equation (\ref{zerilli}) is Zerilli's formula for the $l$-th multipole
component of the energy spectrum in four-dimensions. The total energy spectrum
is given by the sum over the multipoles.

\begin{figure}[h]
\centering
\includegraphics[angle=270,width=9cm,clip]{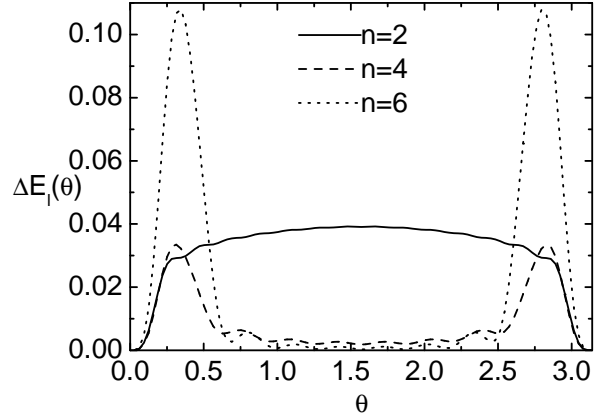}
\caption{
Angular dependence of the radiation for $n=2$, $4$ and $6$, summing all
multipoles up to $l=15$ in units $r_h=1$. The angular distribution for $n=8$ is
not shown. The latter is even more peaked in a narrow region around $\theta=0$
since also the multipoles with $l>15$ contribute significantly to the
radiation.
}
\label{fig4}
\end{figure}

For each $l$ the energy spectrum is
\be
\frac{dE_l}{d\Omega d\omega}=\frac{\omega^2}{32\pi}|\Phi_l|^2
\left(D_iD_j{\cal S}+\frac{\kappa^2}{n}\gamma_{ij}{\cal S}\right)
\left(D^iD^j{\cal S}^*+\frac{\kappa^2}{n}\gamma^{ij}{\cal S}^*\right)\,.
\ee
Substituting Eq.~(\ref{twosided}) in the previous equation, we find
\be
\frac{dE_l}{d\Omega d\omega}=\frac{dE_l}{d\omega}
\frac{n}{n-1}\frac{1}{\kappa^2(\kappa^2-n)}
\left(D_iD_j{\cal S}+\frac{\kappa^2}{n}\gamma_{ij}{\cal S}\right)
\left(D^iD^j{\cal S}^*+\frac{\kappa^2}{n}\gamma^{ij}{\cal S}^*\right)\,.
\ee
The angular dependence for the $l$-th multipole is obtained by integrating
over frequency \cite{Cardoso:2002ay}. The result is
\beq
&&\frac{dE_l}{d\Omega}=\Delta E_l
\frac{n}{n-1}\frac{1}{\kappa^2(\kappa^2-n)}
\left(D_iD_j{\cal S}+\frac{\kappa^2}{n}\gamma_{ij}{\cal S}\right)
\left(D^iD^j{\cal S}^*+\frac{\kappa^2}{n}\gamma^{ij}{\cal S}^*\right)
\equiv\Delta E_l\Lambda_l(\theta)\,,
\eeq
where
\be\label{angular}
\Lambda_l(\theta)=\frac{1}{\kappa^2(\kappa^2-n)}\left(\frac{n}{n-1}
{\cal S}_{,\theta\theta}+\frac{k}{n-1}{\cal S}\right)^2\,.
\ee
For $n=2$ Eq.~(\ref{angular}) reduces to the known result in four dimensions
\cite{Cardoso:2002ay}. The angular dependence is obtained by summing over 
the multipoles:
\be\label{angulartot}
\frac{dE}{d\Omega}(\theta)=\sum_l
\frac{dE_l}{d\Omega}=\sum_l\Delta E_l\Lambda_l(\theta)\,.
\ee
The result truncated to $l_{max}=15$ is shown in the left panel of
Fig.~\ref{fig4}. The curve corresponding to $n=2$ shows good agreement
with Fig.~3 of Ref.~\cite{Cardoso:2002ay}.

\clearpage

\begin{table}
\centering
\caption{
QNMs for $n=2$. The first three quasinormal frequencies for scalar and vector
perturbations are listed from left to right. The scalar modes and the vector
modes are isospectral in four dimensions.
}
\vskip 12pt
\begin{tabular}{@{}|c|c|c|c|@{}}
\hline
n=2
&\multicolumn{3}{|c|}{Scalar and vector modes}\\
\hline
l 	 &$j=0$                &$j=1$             &$j=2$\\
\hline	     
2	 &0.746-0.178\ii~&  0.692-0.550\ii~&  0.606-0.942\ii~\\
3	 &1.199-0.185\ii~&  1.165-0.563\ii~&  1.106-0.953\ii~\\
4	 &1.618-0.188\ii~&  1.593-0.569\ii~&  1.547-0.958\ii~\\
5	 &2.025-0.190\ii~&  2.004-0.572\ii~&  1.967-0.960\ii~\\
6	 &2.424-0.191\ii~&  2.407-0.573\ii~&  2.375-0.961\ii~\\
7	 &2.819-0.191\ii~&  2.805-0.574\ii~&  2.777-0.961\ii~\\
8	 &3.212-0.191\ii~&  3.200-0.575\ii~&  3.175-0.961\ii~\\
9	 &3.604-0.192\ii~&  3.592-0.575\ii~&  3.570-0.962\ii~\\
10	 &3.994-0.192\ii~&  3.983-0.576\ii~&  3.963-0.962\ii~\\
11	 &4.383-0.192\ii~&  4.373-0.576\ii~&  4.355-0.962\ii~\\
12	 &4.771-0.192\ii~&  4.762-0.576\ii~&  4.745-0.962\ii~\\
13	 &5.159-0.192\ii~&  5.151-0.576\ii~&  5.135-0.962\ii~\\
14	 &5.546-0.192\ii~&  5.539-0.577\ii~&  5.524-0.962\ii~\\
15	 &5.934-0.192\ii~&  5.927-0.577\ii~&  5.913-0.962\ii~\\
\hline
\end{tabular}
\label{QNM2}
\end{table}

\begin{table}
\centering
\caption{
QNMs for $n=4$. The first three quasinormal frequencies for scalar, vector, and
tensor perturbations are listed from left to right.
}
\vskip 12pt
\begin{tabular}{@{}|c|c|c|c|c|c|c|c|c|c|@{}}
\hline
n=4
&\multicolumn{3}{|c|}{Scalar modes} 
&\multicolumn{3}{c|}{Vector modes} 
&\multicolumn{3}{c|}{Tensor modes} \\
\hline
l 	 &$j=0$                &$j=1$             &$j=2$             &$j=0$                &$j=1$             &$j=2$            &$j=0$                &$j=1$             &$j=2$\\
\hline								     
2	 & 1.131-0.386\ii~&  0.922-1.186\ii~&  0.537-2.053\ii~&   1.543-0.476\ii~&  1.279-1.482\ii~&  0.825-2.583\ii~&   2.004-0.503\ii~&  1.764-1.568\ii~&  1.378-2.732\ii~\\
3	 & 1.915-0.399\ii~&  1.715-1.217\ii~&  1.336-2.103\ii~&   2.191-0.471\ii~&  1.988-1.445\ii~&  1.625-2.492\ii~&	 2.576-0.499\ii~&  2.393-1.531\ii~&  2.075-2.632\ii~\\
4	 & 2.622-0.438\ii~&  2.476-1.331\ii~&  2.208-2.271\ii~&   2.824-0.474\ii~&  2.664-1.441\ii~&  2.369-2.460\ii~&	 3.146-0.498\ii~&  2.998-1.514\ii~&  2.729-2.580\ii~\\
5	 & 3.279-0.457\ii~&  3.156-1.384\ii~&  2.924-2.347\ii~&   3.441-0.478\ii~&  3.310-1.447\ii~&  3.063-2.453\ii~&	 3.716-0.497\ii~&  3.592-1.504\ii~&  3.359-2.549\ii~\\
6	 & 3.911-0.467\ii~&  3.803-1.412\ii~&  3.598-2.385\ii~&   4.046-0.481\ii~&  3.935-1.453\ii~&  3.723-2.454\ii~&	 4.286-0.496\ii~&  4.179-1.498\ii~&  3.974-2.530\ii~\\
7	 & 4.527-0.474\ii~&  4.432-1.429\ii~&  4.249-2.408\ii~&   4.644-0.484\ii~&  4.547-1.458\ii~&  4.360-2.456\ii~&	 4.856-0.496\ii~&  4.762-1.495\ii~&  4.580-2.517\ii~\\
8	 & 5.133-0.478\ii~&  5.048-1.441\ii~&  4.883-2.422\ii~&   5.236-0.485\ii~&  5.150-1.462\ii~&  4.983-2.458\ii~&	 5.427-0.495\ii~&  5.342-1.492\ii~&  5.178-2.508\ii~\\
9	 & 5.732-0.481\ii~&  5.655-1.449\ii~&  5.505-2.432\ii~&   5.824-0.487\ii~&  5.747-1.466\ii~&  5.596-2.460\ii~&	 5.997-0.495\ii~&  5.921-1.490\ii~&  5.772-2.501\ii~\\
10	 & 6.326-0.484\ii~&  6.256-1.455\ii~&  6.118-2.439\ii~&   6.409-0.488\ii~&  6.339-1.468\ii~&  6.201-2.462\ii~&	 6.567-0.495\ii~&  6.498-1.489\ii~&  6.361-2.496\ii~\\
11	 & 6.916-0.485\ii~&  6.852-1.459\ii~&  6.725-2.444\ii~&   6.992-0.489\ii~&  6.928-1.470\ii~&  6.801-2.463\ii~&	 7.138-0.495\ii~&  7.074-1.488\ii~&  6.948-2.492\ii~\\
12	 & 7.504-0.487\ii~&  7.444-1.463\ii~&  7.326-2.448\ii~&   7.574-0.490\ii~&  7.514-1.472\ii~&  7.396-2.464\ii~&	 7.708-0.495\ii~&  7.649-1.487\ii~&  7.532-2.489\ii~\\
13	 & 8.088-0.488\ii~&  8.033-1.465\ii~&  7.923-2.452\ii~&   8.153-0.490\ii~&  8.098-1.473\ii~&  7.988-2.465\ii~&	 8.279-0.495\ii~&  8.224-1.486\ii~&  8.115-2.487\ii~\\
14	 & 8.671-0.489\ii~&  8.619-1.468\ii~&  8.517-2.454\ii~&   8.732-0.491\ii~&  8.680-1.474\ii~&  8.577-2.465\ii~&	 8.849-0.495\ii~&  8.798-1.486\ii~&  8.696-2.485\ii~\\
15	 & 9.253-0.489\ii~&  9.204-1.469\ii~&  9.108-2.456\ii~&   9.309-0.491\ii~&  9.261-1.475\ii~&  9.164-2.466\ii~&	 9.420-0.495\ii~&  9.371-1.485\ii~&  9.275-2.483\ii~\\
\hline
\end{tabular}
\label{QNM4}
\end{table}

\clearpage

\begin{table}
\centering
\caption{
QNMs for $n=6$. The first three quasinormal frequencies for scalar,
vector, and tensor perturbations are listed from left to right. The
numbers in italic indicate that the potential at the given $l$ is not
everywhere positive definite. The square brackets indicate that the
potential has two scattering peaks.
}
\vskip 12pt
\begin{tabular}{@{}|c|c|c|c|c|c|c|c|c|c|@{}}
\hline
n=4
&\multicolumn{3}{|c|}{Scalar modes} 
&\multicolumn{3}{c|}{Vector modes} 
&\multicolumn{3}{c|}{Tensor modes} \\
\hline
l 	 &$j=0$                &$j=1$             &$j=2$             &$j=0$                &$j=1$             &$j=2$            &$j=0$                &$j=1$             &$j=2$\\
\hline								     
2&{\it [1.778-0.571\ii]}~&{\it [1.289-1.770\ii]}~&{\it[0.395-3.201\ii]}~&{\it 2.388-0.720\ii}~&{\it1.831-2.237\ii}~&{\it 0.825-4.001\ii}~&2.956-0.751\ii~&  2.365-2.357\ii~&  1.339-4.245\ii~\\
3	    & 2.604-0.628\ii~&  2.198-1.916\ii~&  1.403-3.355\ii~ & 3.102-0.715\ii~&  2.660-2.191\ii~&  1.814-3.833\ii~&3.623-0.747\ii~&  3.181-2.294\ii~&  2.351-4.012\ii~\\
4	    & 3.401-0.645\ii~&  3.050-1.958\ii~&  2.346-3.375\ii~ & 3.815-0.712\ii~&  3.450-2.165\ii~&  2.730-3.731\ii~&4.282-0.744\ii~&  3.926-2.264\ii~&  3.235-3.895\ii~\\
5	    & 4.174-0.660\ii~&  3.875-1.997\ii~&  3.270-3.403\ii~ & 4.522-0.712\ii~&  4.213-2.156\ii~&  3.595-3.678\ii~&4.940-0.741\ii~&  4.640-2.247\ii~&  4.047-3.830\ii~\\
6	    & 4.923-0.675\ii~&  4.665-2.037\ii~&  4.144-3.449\ii~ & 5.222-0.714\ii~&  4.954-2.156\ii~&  4.418-3.654\ii~&5.598-0.740\ii~&  5.337-2.236\ii~&  4.818-3.789\ii~\\
7	    & 5.653-0.687\ii~&  5.425-2.070\ii~&  4.967-3.492\ii~ & 5.915-0.716\ii~&  5.679-2.160\ii~&  5.207-3.645\ii~&6.255-0.739\ii~&  6.024-2.229\ii~&  5.563-3.763\ii~\\
8	    & 6.369-0.695\ii~&  6.164-2.095\ii~&  5.753-3.525\ii~ & 6.602-0.719\ii~&  6.392-2.164\ii~&  5.969-3.643\ii~&6.913-0.738\ii~&  6.705-2.224\ii~&  6.290-3.745\ii~\\
9	    & 7.075-0.702\ii~&  6.888-2.113\ii~&  6.515-3.550\ii~ & 7.285-0.721\ii~&  7.094-2.169\ii~&  6.712-3.644\ii~&7.570-0.738\ii~&  7.382-2.221\ii~&  7.004-3.732\ii~\\
10	    & 7.772-0.707\ii~&  7.602-2.128\ii~&  7.259-3.570\ii~ & 7.964-0.722\ii~&  7.790-2.173\ii~&  7.441-3.646\ii~&8.228-0.737\ii~&  8.055-2.218\ii~&  7.709-3.722\ii~\\
11	    & 8.464-0.711\ii~&  8.306-2.139\ii~&  7.989-3.585\ii~ & 8.640-0.724\ii~&  8.480-2.177\ii~&  8.158-3.648\ii~&8.885-0.737\ii~&  8.726-2.216\ii~&  8.406-3.715\ii~\\
12	    & 9.151-0.715\ii~&  9.004-2.148\ii~&  8.709-3.598\ii~ & 9.314-0.725\ii~&  9.165-2.180\ii~&  8.867-3.650\ii~&9.543-0.737\ii~&  9.395-2.215\ii~&  9.098-3.709\ii~\\
13	    & 9.834-0.717\ii~&  9.697-2.156\ii~&  9.421-3.607\ii~ & 9.986-0.726\ii~&  9.847-2.183\ii~&  9.569-3.653\ii~&10.200-0.737\ii~&  10.062-2.214\ii~&  9.785-3.705\ii~\\
14	    & 10.51-0.720\ii~&  10.39-2.162\ii~&  10.13-3.616\ii~ & 10.66-0.727\ii~&  10.53-2.185\ii~&  10.27-3.655\ii~&10.86-0.737\ii~&  10.73-2.213\ii~&  10.47-3.701\ii~\\
15	    & 11.19-0.721\ii~&  11.07-2.167\ii~&  10.83-3.622\ii~ & 11.32-0.728\ii~&  11.20-2.187\ii~&  10.96-3.657\ii~&11.52-0.736\ii~&  11.39-2.212\ii~&  11.15-3.698\ii~\\
\hline
\end{tabular}
\label{QNM6}
\end{table}

\begin{table}
\centering
\caption{
QNMs for $n=8$. The first three quasinormal frequencies for scalar,
vector, and tensor perturbations are listed from left to right. The
numbers in italic indicate that the potential at the given $l$ is not
everywhere positive definite. The square brackets indicate that the
potential has two scattering peaks.
}
\vskip 12pt
\begin{tabular}{@{}|c|c|c|c|c|c|c|c|c|c|@{}}
\hline
n=4
&\multicolumn{3}{|c|}{Scalar modes} 
&\multicolumn{3}{c|}{Vector modes} 
&\multicolumn{3}{c|}{Tensor modes} \\
\hline
l 	 &$j=0$                &$j=1$             &$j=2$             &$j=0$                &$j=1$             &$j=2$            &$j=0$                &$j=1$             &$j=2$\\
\hline								     
l = 2 &{\it [2.513-0.744\ii]}~&{\it [1.686-2.299\ii]}~&{\it [0.159-4.345\ii]}~&{\it 3.261-0.924\ii}~&{\it 2.335-2.851\ii}~&{\it 0.598-5.287\ii}~&3.886-0.959\ii~&  2.765-2.988\ii~&  0.706-5.720\ii~\\
l = 3 &[3.388-0.812\ii]~  &[2.696-2.461\ii]~  &[1.277-4.431\ii]~&{\it 4.017-0.923\ii}~&{\it 3.269-2.804\ii}~&{\it 1.747-5.016\ii}~&4.618-0.959\ii~&  3.806-2.917\ii~&  2.141-5.241\ii~\\
l = 4	 & 4.223-0.841\ii~&  3.631-2.532\ii~&  2.367-4.420\ii~&     4.775-0.920\ii~&  4.147-2.777\ii~&  2.824-4.840\ii~& 5.336-0.955\ii~&  4.691-2.885\ii~&  3.331-5.018\ii~\\
l = 5	 & 5.042-0.855\ii~&  4.524-2.568\ii~&  3.407-4.401\ii~&     5.531-0.918\ii~&  4.991-2.762\ii~&  3.840-4.734\ii~& 6.049-0.951\ii~&  5.507-2.866\ii~&  4.360-4.904\ii~\\
l = 6	 & 5.848-0.865\ii~&  5.390-2.595\ii~&  4.403-4.399\ii~&     6.283-0.917\ii~&  5.810-2.757\ii~&  4.802-4.676\ii~& 6.761-0.949\ii~&  6.291-2.854\ii~&  5.297-4.838\ii~\\
l = 7	 & 6.640-0.874\ii~&  6.231-2.622\ii~&  5.357-4.415\ii~&     7.030-0.918\ii~&  6.610-2.757\ii~&  5.719-4.646\ii~& 7.473-0.947\ii~&  7.056-2.846\ii~&  6.178-4.798\ii~\\
l = 8	 & 7.420-0.883\ii~&  7.052-2.647\ii~&  6.270-4.441\ii~&     7.774-0.920\ii~&  7.396-2.759\ii~&  6.599-4.634\ii~& 8.184-0.946\ii~&  7.808-2.841\ii~&  7.022-4.772\ii~\\
l = 9	 & 8.191-0.890\ii~&  7.855-2.669\ii~&  7.149-4.469\ii~&     8.514-0.921\ii~&  8.167-2.764\ii~&  7.450-4.630\ii~& 8.895-0.945\ii~&  8.553-2.837\ii~&  7.841-4.755\ii~\\
l = 10	 & 8.953-0.896\ii~&  8.645-2.688\ii~&  7.999-4.495\ii~&     9.250-0.923\ii~&  8.935-2.768\ii~&  8.278-4.630\ii~& 9.606-0.944\ii~&  9.292-2.834\ii~&  8.640-4.743\ii~\\
l = 11	 & 9.709-0.902\ii~&  9.423-2.705\ii~&  8.829-4.519\ii~&     9.984-0.924\ii~&  9.692-2.773\ii~&  9.088-4.633\ii~& 10.32-0.944\ii~&  10.03-2.832\ii~&  9.426-4.735\ii~\\
l = 12	 & 10.46-0.906\ii~&  10.19-2.718\ii~&  9.641-4.539\ii~&     10.71-0.926\ii~&  10.44-2.777\ii~&  9.884-4.637\ii~& 11.03-0.943\ii~&  10.76-2.831\ii~&  10.20-4.728\ii~\\
l = 13	 & 11.20-0.910\ii~&  10.95-2.730\ii~&  10.44-4.556\ii~&     11.44-0.927\ii~&  11.19-2.781\ii~&  10.67-4.642\ii~& 11.74-0.943\ii~&  11.49-2.829\ii~&  10.97-4.724\ii~\\
l = 14	 & 11.95-0.914\ii~&  11.71-2.740\ii~&  11.23-4.571\ii~&     12.17-0.928\ii~&  11.93-2.785\ii~&  11.44-4.646\ii~& 12.45-0.943\ii~&  12.21-2.828\ii~&  11.73-4.720\ii~\\
l = 15	 & 12.68-0.916\ii~&  12.46-2.748\ii~&  12.01-4.584\ii~&     12.90-0.929\ii~&  12.67-2.788\ii~&  12.21-4.650\ii~& 13.16-0.943\ii~&  12.94-2.828\ii~&  12.48-4.718\ii~\\
\hline
\end{tabular}
\label{QNM8}
\end{table}

\clearpage

\begin{table}
\centering
\caption{
Multipolar contributions to the total energy for different $n$ in
units $r_h=1$.
}
\vskip 12pt
\begin{tabular}{@{}|c|c|c|c|c|@{}}
\hline
$l$ &$n=2$    &$n=4$    &$n=6$    &$n=8$\\
\hline
2   &0.1845  &0.189e-1  &0.194e-2   &0.269e-3\\
3   &0.0855  &0.120e-1  &0.238e-2   &0.653e-3\\
4   &0.0500  &0.086e-1  &0.241e-2   &0.983e-3\\
5   &0.0329  &0.064e-1  &0.224e-2   &1.187e-3\\
6   &0.0234  &0.050e-1  &0.199e-2   &1.258e-3\\
7   &0.0175  &0.039e-1  &0.172e-2   &1.225e-3\\
8   &0.0136  &0.032e-1  &0.149e-2   &1.130e-3\\
9   &0.0109  &0.027e-1  &0.128e-2   &1.009e-3\\
10  &0.0089  &0.022e-1  &0.111e-2   &0.888e-3\\
11  &0.0074  &0.019e-1  &0.096e-2   &0.780e-3\\
12  &0.0063  &0.016e-1  &0.084e-2   &0.688e-3\\
13  &0.0054  &0.014e-1  &0.075e-2   &0.612e-3\\
14  &0.0047  &0.013e-1  &0.066e-2   &0.549e-3\\
15  &0.0041  &0.011e-1  &0.059e-2   &0.497e-3\\
\hline			      		  
\end{tabular}
\label{El}
\end{table}

\begin{table}
\centering
\caption{
Total energy for different spacetime dimensions. From left to right, the
columns give the spacetime dimension $D=n+2$, the factor $n {\cal A}_n/16\pi$,
the total energy $E^{(D)}_{\rm tot}$ in units $r_h=1$,  the rescaled total
energy  ${\cal E}_{\rm tot}^{(D)}=n{\cal A}_n E^{(D)}_{\rm tot}/16\pi$  and the
gravitational energy loss (see text).
}
\vskip 12pt
\begin{tabular}{@{}|c|c|c|c|c|@{}}
\hline
$D$ &$n {\cal A}_n/16\pi$  &$E^{(D)}_{\rm tot}$ &${\cal E}_{\rm tot}^{(D)}$
&Energy loss \\
\hline
4   &$1/2$  		&0.52   &0.26   &13\%\\
6   &$2\pi/3$  		&0.095  &0.20   &10\%\\
8   &$2\pi^2/5$  	&0.034  &0.13   &7\%\\
10  &$16\pi^3/105$  	&0.032  &0.15   &8\%\\
\hline			      		  
\end{tabular}
\label{Etot}
\end{table}


\thebibliography{99}
%
\bibitem{extradim}
I.~Antoniadis,
Phys.\ Lett.\ B {\bf 246}, 377 (1990);
N.~Arkani-Hamed, S.~Dimopoulos and G.~R.~Dvali,
Phys.\ Lett.\ B {\bf 429}, 263 (1998)
[arXiv:hep-ph/9803315];
I.~Antoniadis, N.~Arkani-Hamed, S.~Dimopoulos and G.~R.~Dvali,
Phys.\ Lett.\ B {\bf 436}, 257 (1998)
[arXiv:hep-ph/9804398];
L.~Randall and R.~Sundrum,
Phys.\ Rev.\ Lett.\  {\bf 83}, 3370 (1999)
[arXiv:hep-ph/9905221];
L.~Randall and R.~Sundrum,
Phys.\ Rev.\ Lett.\  {\bf 83}, 4690 (1999)
[arXiv:hep-th/9906064].

\bibitem{pertextradim}
G.~F.~Giudice, R.~Rattazzi and J.~D.~Wells,
Nucl.\ Phys.\ B {\bf 544}, 3 (1999)
[arXiv:hep-ph/9811291];
T.~Han, J.~D.~Lykken and R.~J.~Zhang,
Phys.\ Rev.\ D {\bf 59}, 105006 (1999)
[arXiv:hep-ph/9811350];
J.~L.~Hewett,
Phys.\ Rev.\ Lett.\  {\bf 82}, 4765 (1999)
[arXiv:hep-ph/9811356];
T.~G.~Rizzo,
Phys.\ Rev.\ D {\bf 59}, 115010 (1999)
[arXiv:hep-ph/9901209].

\bibitem{nonpertextradim}
T.~Banks and W.~Fischler,
arXiv:hep-th/9906038;
S.~B.~Giddings and S.~Thomas,
Phys.\ Rev.\ D {\bf 65}, 056010 (2002)
[arXiv:hep-ph/0106219];
S.~Dimopoulos and G.~Landsberg,
Phys.\ Rev.\ Lett.\  {\bf 87}, 161602 (2001)
[arXiv:hep-ph/0106295];
S.~Dimopoulos and R.~Emparan,
Phys.\ Lett.\ B {\bf 526}, 393 (2002)
[arXiv:hep-ph/0108060];
R.~Casadio and B.~Harms,
Int.\ J.\ Mod.\ Phys.\ A {\bf 17}, 4635 (2002)
[arXiv:hep-th/0110255];
E.~J.~Ahn, M.~Cavagli\`a and A.~V.~Olinto,
Phys.\ Lett.\ B {\bf 551}, 1 (2003)
[arXiv:hep-th/0201042].

\bibitem{uhecr}
J.~L.~Feng and A.~D.~Shapere,
Phys.\ Rev.\ Lett.\  {\bf 88}, 021303 (2002)
[arXiv:hep-ph/0109106];
A.~Ringwald and H.~Tu,
Phys.\ Lett.\ B {\bf 525}, 135 (2002)
[arXiv:hep-ph/0111042];
E.~J.~Ahn, M.~Ave, M.~Cavagli\`a and A.~V.~Olinto,
Phys.\ Rev.\ D {\bf 68}, 043004 (2003)
[arXiv:hep-ph/0306008].

\bibitem{hardscat}
S.~Hossenfelder, S.~Hofmann, M.~Bleicher and H.~Stocker,
Phys.\ Rev.\ D {\bf 66}, 101502 (2002)
[arXiv:hep-ph/0109085];
K.~Cheung,
Phys.\ Rev.\ Lett.\  {\bf 88}, 221602 (2002)
[arXiv:hep-ph/0110163];
S.~I.~Dutta, M.~H.~Reno and I.~Sarcevic,
Phys.\ Rev.\ D {\bf 66}, 033002 (2002)
[arXiv:hep-ph/0204218];
K.~Cheung,
Phys.\ Rev.\ D {\bf 66}, 036007 (2002)
[arXiv:hep-ph/0205033];
E.~J.~Ahn and M.~Cavagli\`a,
Gen.\ Rel.\ Grav.\  {\bf 34}, 2037 (2002)
[arXiv:hep-ph/0205168].;
V.~Frolov and D.~Stojkovic,
Phys.\ Rev.\ D {\bf 66}, 084002 (2002)
[arXiv:hep-th/0206046];
T.~Han, G.~D.~Kribs and B.~McElrath,
Phys.\ Rev.\ Lett.\  {\bf 90}, 031601 (2003)
[arXiv:hep-ph/0207003];
V.~Frolov and D.~Stojkovic,
Phys.\ Rev.\ Lett.\  {\bf 89}, 151302 (2002)
[arXiv:hep-th/0208102];
I.~Mocioiu, Y.~Nara and I.~Sarcevic,
Phys.\ Lett.\ B {\bf 557}, 87 (2003)
[arXiv:hep-ph/0301073];
A.~Chamblin, F.~Cooper and G.~C.~Nayak,
arXiv:hep-ph/0301239.

\bibitem{Cavaglia:2002si}
M.~Cavagli\`a,
Int.\ J.\ Mod.\ Phys.\ A {\bf 18}, 1843 (2003)
[arXiv:hep-ph/0210296].

\bibitem{hoop}
K.S.~Thorne,
in: {\it Magic without magic: John Archibald Wheeler},
edited by J.~Klauder (freeman, San Francisco, 1972).

\bibitem{relic}
M.~Cavagli\`a, S.~Das and R.~Maartens,
Class.\ Quant.\ Grav.\  {\bf 20}, L205 (2003)
[arXiv:hep-ph/0305223];
S.~Hossenfelder, M.~Bleicher, S.~Hofmann, H.~Stocker and A.~V.~Kotwal,
arXiv:hep-ph/0302247.

\bibitem{Cavaglia:2003hg}
M.~Cavagli\`a,
Phys.\ Lett.\ B {\bf 569}, 7 (2003)
[arXiv:hep-ph/0305256].

\bibitem{Yoshino:2002tx}
H.~Yoshino and Y.~Nambu,
Phys.\ Rev.\ D {\bf 67}, 024009 (2003)
[arXiv:gr-qc/0209003].

\bibitem{Vasilenko:2003ak}
O.~I.~Vasilenko,
arXiv:hep-th/0305067.

\bibitem{apphor}
P.~D.~D'Eath and P.~N.~Payne,
Phys.\ Rev.\ D {\bf 46}, 658 (1992);
P.~D.~D'Eath and P.~N.~Payne,
Phys.\ Rev.\ D {\bf 46}, 675 (1992);
P.~D.~D'Eath and P.~N.~Payne,
Phys.\ Rev.\ D {\bf 46}, 694 (1992);
D.~M.~Eardley and S.~B.~Giddings,
Phys.\ Rev.\ D {\bf 66}, 044011 (2002)
[arXiv:gr-qc/0201034];
H.~Yoshino and Y.~Nambu,
Phys.\ Rev.\ D {\bf 66}, 065004 (2002)
[arXiv:gr-qc/0204060];
D.~Ida and K.~I.~Nakao,
Phys.\ Rev.\ D {\bf 66}, 064026 (2002)
[arXiv:gr-qc/0204082].
O.~I.~Vasilenko,
arXiv:hep-th/0305067.

\bibitem{Aichelburg:1970dh}
P.~C.~Aichelburg and R.~U.~Sexl,
Gen.\ Rel.\ Grav.\  {\bf 2}, 303 (1971).

\bibitem{LHC}
http://lhc-new-homepage.web.cern.ch/lhc-new-homepage/

\bibitem{Gleiser:1998rw}
R.~J.~Gleiser, C.~O.~Nicasio, R.~H.~Price and J.~Pullin,
Phys.\ Rept.\  {\bf 325}, 41 (2000)
[arXiv:gr-qc/9807077].

\bibitem{Anninos:1998wt}
P.~Anninos and S.~Brandt,
Phys.\ Rev.\ Lett.\  {\bf 81}, 508 (1998)
[arXiv:gr-qc/9806031].

\bibitem{RW}
T.~Regge, J.~A.~Wheeler,
Phys. Rev. {\bf 108}, 1063 (1957).

\bibitem{Zerilli}
F.~Zerilli,
Phys. Rev. D {\bf 2}, 2141 (1970).

\bibitem{DRPP}
M.~Davis, R.~Ruffini, W.~H.~Press, R.~H.~Price,
Phys. Rev. Lett. {\bf 27}, 1466 (1971).

\bibitem{Ruffini}
R.~Ruffini,
Phys. Rev. D {\bf 7}, 972 (1973);
R.~Ruffini, V.~Ferrari, 
Phys. Lett. B {\bf 98}, 381 (1981);
C.~Lousto, R.~H.~Price,
Phys. Rev. D {\bf 55}, 2124 (1997).

\bibitem{perturb}
J.~Baker, B.~Brugmann, M.~Campanelli, C.~O.~Lousto and R.~Takahashi,
Phys.\ Rev.\ Lett.\  {\bf 87}, 121103 (2001)
[arXiv:gr-qc/0102037];
Y.~Mino, M.~Sasaki, M.~Shibata, H.~Tagoshi and T.~Tanaka,
Prog.\ Theor.\ Phys.\ Suppl.\  {\bf 128}, 1 (1997)
[arXiv:gr-qc/9712057];
S.~Chandrasekhar, {\em The Mathematical Theory of Black Holes}, 
Oxford University Press (1983).

\bibitem{Cardoso:2002ay}
V.~Cardoso and J.~P.~Lemos,
Phys.\ Lett.\ B {\bf 538}, 1 (2002)
[arXiv:gr-qc/0202019].

\bibitem{Cardoso:2002yj}
V.~Cardoso and J.~P.~Lemos,
Gen.\ Rel.\ Grav.\  {\bf 35}, 327 (2003)
[arXiv:gr-qc/0207009].

\bibitem{Smarr:1977}
L.~Smarr,
Phys. Rev. D {\bf 15}, 2069 (1977).

\bibitem{Cardoso:2002pa}
V.~Cardoso, O.~J.~Dias and J.~P.~Lemos,
Phys.\ Rev.\ D {\bf 67}, 064026 (2003)
[arXiv:hep-th/0212168].

\bibitem{Cardoso:2003jf}
V.~Cardoso, S.~Yoshida, O.~J.~Dias and J.~P.~Lemos,
arXiv:hep-th/0307122.

\bibitem{Barvinsky:2003jf}
A.~O.~Barvinsky and S.~N.~Solodukhin,
arXiv:hep-th/0307011.

\bibitem{Hollands:2003ie}
S.~Hollands and A.~Ishibashi,
arXiv:gr-qc/0304054.

\bibitem{Kodama:2000fa}
H.~Kodama, A.~Ishibashi and O.~Seto,
Phys.\ Rev.\ D {\bf 62}, 064022 (2000)
[arXiv:hep-th/0004160].

\bibitem{Kodama:2003jz}
H.~Kodama and A.~Ishibashi,
arXiv:hep-th/0305147.

\bibitem{Ishibashi:2003ap}
A.~Ishibashi and H.~Kodama,
arXiv:hep-th/0305185.

\bibitem{Kodama:2003kk}
H.~Kodama and A.~Ishibashi,
arXiv:hep-th/0308128.

\bibitem{Konoplya:2003ii}
R.~A.~Konoplya,
Phys.\ Rev.\ D {\bf 68}, 024018 (2003)
[arXiv:gr-qc/0303052].

\bibitem{Konoplya:2003dd}
R.~A.~Konoplya,
Phys.\ Rev.\ D {\bf 68}, 124017 (2003)
[arXiv:hep-th/0309030].

\bibitem{Tangherlini:1963}
F.~R.~Tangherlini,
Nuovo Cimento {\bf 27}, 636 (1963).

\bibitem{Gibbons:2002pq}
G.~Gibbons and S.~A.~Hartnoll,
Phys.\ Rev.\ D {\bf 66}, 064024 (2002)
[arXiv:hep-th/0206202].

\bibitem{Neupane}
I. P. Neupane,
arXiv:hep-th/0302132.

\bibitem{KS}
K.~D.~Kokkotas and B.~G.~Schmidt,
Living Rev.\ Rel.\  {\bf 2}, 2 (1999)
[arXiv:gr-qc/9909058];
H.-P. Nollert, 
Class.\ Quant.\ Grav.\ {\bf 16}, R159 (1999).

\bibitem{Iyer:np}
S.~Iyer and C.~M.~Will,
Phys.\ Rev.\ D {\bf 35}, 3621 (1987).

\bibitem{Iyer:nq}
S.~Iyer,
Phys.\ Rev.\ D {\bf 35}, 3632 (1987).

\bibitem{Hod:1998vk}
S.~Hod,
Phys.\ Rev.\ Lett.\  {\bf 81}, 4293 (1998)
[arXiv:gr-qc/9812002].

\bibitem{Motl:2003cd}
L.~Motl and A.~Neitzke,
arXiv:hep-th/0301173.

\bibitem{Birmingham:2003rf}
D.~Birmingham,
arXiv:hep-th/0306004.

\bibitem{AreaQuantum}
O.~Dreyer,
Phys.\ Rev.\ Lett.\  {\bf 90}, 081301 (2003)
[arXiv:gr-qc/0211076];
G.~Kunstatter,
Phys.\ Rev.\ Lett.\  {\bf 90}, 161301 (2003)
[arXiv:gr-qc/0212014];
L.~Motl,
Adv.\ Theor.\ Math.\ Phys.\  {\bf 6}, 1135 (2003)
[arXiv:gr-qc/0212096];
E.~Berti and K.~D.~Kokkotas,
Phys.\ Rev.\ D {\bf 67}, 064020 (2003)
[arXiv:gr-qc/0301052];
V.~Cardoso and J.~P.~Lemos,
Phys.\ Rev.\ D {\bf 67}, 084020 (2003)
[arXiv:gr-qc/0301078];
C.~Molina,
Phys.\ Rev.\ D {\bf 68}, 064007 (2003)
[arXiv:gr-qc/0304053];
A.~Neitzke,
arXiv:hep-th/0304080;
A.~Maassen van den Brink,
Phys.\ Rev.\ D {\bf 68}, 047501 (2003)
[arXiv:gr-qc/0304092];
V.~Cardoso, R.~Konoplya and J.~P.~Lemos,
Phys.\ Rev.\ D {\bf 68}, 044024 (2003)
[arXiv:gr-qc/0305037];
D.~Birmingham, S.~Carlip and Y.~j.~Chen,
arXiv:hep-th/0305113;
E.~Berti, V.~Cardoso, K.~D.~Kokkotas and H.~Onozawa,
arXiv:hep-th/0307013;
N.~Andersson and C.~J.~Howls,
arXiv:gr-qc/0307020;
S.~Hod,
arXiv:gr-qc/0307060;
S.~Yoshida and T.~Futamase,
arXiv:gr-qc/0308077;
S.~Musiri and G.~Siopsis,
arXiv:hep-th/0308168.
J.~Oppenheim,
arXiv:gr-qc/0307089.

\bibitem{Weinberg}
S.~Weinberg,
{\it Gravitation and Cosmology} 
(Wiley, New York, 1972).

\bibitem{Cardoso:2002cf}
V.~Cardoso and J.~P.~Lemos,
Phys.\ Rev.\ D {\bf 66}, 064006 (2002)
[arXiv:hep-th/0206084].

\bibitem{Abr} 
M.~Abramowitz, I.~A.~Stegun, 
in {\it Handbook of Mathematical functions} (Dover, New York, 1970); 
useful information can also be found at {\tt http://www.mathworld.wolfram.com}.
\end{document}